\documentclass[useAMS,usenatbib]{mnras}
\usepackage{graphicx,amssymb,amsmath,times,verbatim,pdfpages,multirow,booktabs,array,dcolumn}
\usepackage[caption=false]{subfig}
\usepackage[flushleft]{threeparttable}
\usepackage[countmax]{subfloat}
\usepackage{xcolor}
\usepackage{ulem}
\usepackage{soul}

\title[Optical Variability of PG 1553$+$113]{Multi-band Optical Variability of the TeV Blazar PG 1553$+$113 in 2019}
\author[Dhiman et al.]{Vinit Dhiman$^{1,2}$\thanks{Email: dhiman@aries.res.in}, Alok C. Gupta$^{1,3}$\thanks{Email: acgupta30@gmail.com}, Sofia O. Kurtanidze$^{4,5}$, I. Eglitis$^{6}$, A. Strigachev$^{7}$, G. Damljanovic$^{8}$,
\newauthor Paul J. Wiita$^{9}$, Minfeng Gu$^{3}$, Haritma Gaur$^{1}$, Oliver Vince$^{8}$, R. Bachev$^{7}$, D. P. Bisen$^{2}$, S. Ibryamov$^{10}$, 
\newauthor R. Z. Ivanidze$^{4}$, Miljana D. Jovanovic$^{8}$, Omar M. Kurtanidze$^{4,11,5}$, M. G. Nikolashvili$^{4,5}$, E. Semkov$^{7}$, 
\newauthor B. Spassov$^{7}$, M. Stojanovic$^{8}$, Beatriz Villarroel$^{12,13}$, Haiguang Xu$^{14,15}$, Zhongli Zhang$^{16,17,14}$ \\
\\
$^{1}$Aryabhatta Research Institute of Observational Sciences (ARIES), Manora Peak, Nainital 263001, India \\
$^{2}$School of Studies in Physics \& Astrophysics, Pt.\ Ravishankar Shukla University, Amanaka G.E. Road, Raipur 492010, India \\
$^{3}$Key Laboratory for Research in Galaxies and Cosmology, Shanghai Astronomical Observatory, Chinese Academy of Sciences, Shanghai 200030, China \\
$^{4}$Abastumani Observatory, Mt. Kanobili, 0301 Abastumani, Georgia \\
$^{5}$Landessternwarte, Zentrum f$\ddot{u}$r Astronomie der Universit$\ddot{a}$t Heidelberg, K$\ddot{o}$nigstuhl 12, 69117 Heidelberg, Germany \\ 
$^{6}$Institute of Astronomy, University of Latvia, Jelgavas 3, Riga, LV-1004, Latvia \\
$^{7}$Institute of Astronomy and National Astronomical Observatory, Bulgarian Academy of Sciences, 72 Tsarigradsko Shosse Blvd., 1784 Sofia, Bulgaria \\
$^{8}$Astronomical Observatory, Volgina 7, 11060 Belgrade, Serbia \\
$^{9}$Department of Physics, The College of New Jersey, 2000 Pennington Rd., Ewing, NJ 08628-0718, USA \\
$^{10}$Department of Physics and Astronomy, Faculty of Natural Sciences, University of Shumen, Shumen, Bulgaria \\
$^{11}$Engelhardt Astronomical Observatory, Kazan Federal University, Tatarstan, Russia \\
$^{12}$Nordita, KTH Royal Institute of Technology and Stockholm University, Roslagstullsbacken 23, SE-106 91 Stockholm, Sweden \\
$^{13}$Instituto de Astrofisica de Canarias, Avda Via Lactea S/N, La Laguna, E-38205, Tenerife, Spain \\
$^{14}$Shanghai Frontiers Science Center of Gravitational Wave Detection, 800 Dongchuan Road, Minhang, Shanghai 200240, China \\
$^{15}$School of Physics and Astronomy, Shanghai Jiao Tong University, 800 Dongchuan Road, Minhang, Shanghai 200240, China \\
$^{16}$SKA Regional Centre Joint Lab, Shanghai Astronomical Observatory, Chinese Academy of Sciences, Shanghai, 200030, China \\
$^{17}$Key Laboratory of Radio Astronomy, Chinese Academy of Sciences, 210033 Nanjing, Jiangsu, China}

\begin{document}
\label{firstpage}
\pagerange{\pageref{firstpage}--\pageref{lastpage}}
\maketitle

\begin{abstract}
\noindent
We report the flux and spectral variability of PG 1553$+$113 on intra-night (IDV) to short-term  timescales using BVRI data collected over 91 nights from 28 February to 8 November 2019 employing ten optical telescopes: three in Bulgaria, two each in India and Serbia, and one each in Greece, Georgia, and Latvia. We monitored the blazar quasi-simultaneously for 16 nights in the V and R bands and 8 nights in the V, R, I bands and examined the light curves (LCs) for intra-day flux and colour variations using two powerful tests: the power-enhanced F-test and the nested ANOVA test. The source was found to be significantly ($>99\%$) variable in 4 nights out of 27 in R-band, 1 out of 16 in V-band, and 1 out of 6 nights in I-band. No temporal variations in the colours were observed on IDV timescale. During the  course of these observations the total variation in R-band was 0.89 mag observed. We also investigated the spectral energy distribution (SED) using B, V, R, and I band data. We found optical spectral indices in the range of 0.878$\pm$0.029 to 1.106$\pm$0.065 by fitting a power law ($F_{\nu} \propto \nu^{-\alpha}$) to these SEDs of PG 1553$+$113. We found that the source follows a bluer-when-brighter trend on IDV timescales. We discuss possible physical causes of the observed spectral variability.

\end{abstract}


\section{Introduction} \label{sec:introdues}
\noindent
The blazar sub-class of radio-loud (RL) active galactic nuclei (AGN) possess a relativistic jet aligned at an angle of $\leq 10^{\circ}$ from the observer’s line of sight \citep{1995PASP..107..803U}. The Doppler enhanced intense non-thermal radiation from this jet dominates the spectral energy distribution (SED) from radio to very high energy (VHE) $\gamma$-ray energies. Blazars are usually considered to be comprised of both of BL Lacertae (BL Lac) objects, and flat spectrum radio quasars (FSRQs). Blazars show flux and spectral variability across the entire electromagnetic (EM) spectrum, emit predominantly non-thermal radiation showing strong polarization from radio to optical ($>3\%$) frequencies, and usually have core dominated radio structures. \\
\\
The multi-wavelength (MW) SEDs of blazars in the usual log($\nu F_{\nu}$) vs log($\nu$) representation show double-humped structures in which the low energy hump peaks in infrared (IR) through X-ray bands while the  high energy hump peaks in $\gamma$-rays energies \citep{1998MNRAS.299..433F}. The location of SED peaks are often used to classify blazars into two sub-classes namely LBLs (low-energy-peaked blazars) and HBLs (high-energy-peaked blazars). In LBLs, the first hump peaks in IR to optical bands and the second hump peaks at GeV $\gamma$-ray energies. In HBLs, the first hump peaks in UV to X-ray bands and the second hump is located up to TeV $\gamma$-ray energies \citep{1995MNRAS.277.1477P}. The emission of the lower energy SED is due to synchrotron radiation which originates from relativistic electrons in the jet but the high energy portion of the SED can arise in several ways. \\
\\
Flux variability over a wide range of timescales is one of the definitional properties of blazars. On the basis of the times over which it is observed, blazar variability can be divided into three classes: microvariability \citep{1989Natur.337..627M} or intra-day variability (IDV) \citep{1995ARA&A..33..163W} or intra-night variability  \citep{1993MNRAS.262..963G} (occurring on a timescale of a few minutes to less than a day); short-term variability (STV; taking place on a timescale of days to months); and long-term variability (LTV; over a timescale of several months to years or even decades \citep{2004A&A...422..505G}. The first clear optical IDV detection was reported in the light curves of the blazar BL Lacertae \citep{1989Natur.337..627M}. Since then the optical variability of blazars on diverse timescales have been studied extensively and reported in many series of papers \citep[e.g.,][and references therein]{1989Natur.337..627M,1990AJ....100..347C,1992AJ....104...15C,1992ApJS...80..683X,1994A&AS..106..361X,2002MNRAS.329..689X,1996A&A...305...42H,1996A&A...305L..17S,1996A&A...315L..13S,1997A&AS..125..525F,1998ApJ...507..173F,2001A&A...369..758F,1998A&AS..127..445R,1998A&AS..132...83B,2005MNRAS.356..607S,2006A&A...450...39G,2008AJ....135.1384G,2016MNRAS.458.1127G,2017MNRAS.465.4423G,2019AJ....157...95G,2012MNRAS.424.2625B,2012MNRAS.420.3147G,2012MNRAS.425.3002G,2015MNRAS.452.4263G,2019MNRAS.484.5633G,2015MNRAS.450..541A,2016MNRAS.455..680A,2019MNRAS.488.4093A,2019ApJ...871..192P,2020ApJ...890...72P,2020MNRAS.496.1430P}. In the optical regime, HBLs are found to be less variable than LBLs and their variability amplitudes are also much smaller than that of LBLs \citep{1994ApJ...428..130J,2012AJ....143...23G,2012MNRAS.420.3147G}. Measurements of variability amplitudes and duty cycles, temporal lags between bands, along with spectral changes, can provide information about the location, size, structure, and dynamics of the regions emitting non-thermal photons.\\
\\
PG 1553$+$113 was discovered by the Palomar-Green survey of UV-excess stellar objects \citep{1986ApJS...61..305G} and was classified as a BL Lac object due to its featureless optical spectrum and significant optical variability \citep{1983BAAS...15..957M}. Furthermore, it was classified as an HBL, as its synchrotron emission peak falls in the UV and X-ray frequency ranges \citep{1990PASP..102.1120F} and it was detected at TeV energies \citep{2006A&A...448L..19A}. The recent detection of its putative galaxy group would set this object’s redshift at $z = 0.433$ \citep{2019ApJ...884L..31J}. Being relatively bright and with intriguing flux variability, over the past two decades PG 1553$+$113 has been extensively studied in single EM bands in isolation \citep[e.g.,][]{2021A&A...645A.137A,2019ApJ...871..192P,2020MNRAS.492.1295P} or in simultaneous / quasi-simultaneous MW observations on diverse timescales \citep[e.g.,][]{2015ApJ...813L..41A,2021MNRAS.506.1198D}. An optical flare was observed in April 2019 when it showed the brightest magnitude of 13.2 in the R-band over the period 2005 -- 2019 \citep{2021A&A...645A.137A}. On IDV time scales, it has shown significant variation in linear polarization percentage and position angle \citep{2011A&A...531A..38A}.\\
\\
PG 1553$+$113 is among a few blazars which have been claimed to show  occasional periodic / quasi periodic oscillations (QPO) in the light curves in different EM bands on diverse timescales. The first claim of 2.18$\pm$0.08 years periodicity was reported in the $\gamma$-ray, optical and radio light curve (LCs) of the source \citep{2015ApJ...813L..41A}. This period in $\gamma$-rays was confirmed by \citep{2017MNRAS.471.3036P,2018ApJ...854...11T,2018A&A...615A.118S,2020ApJ...895..122C,2020ApJ...896..134P} and also confirmed in the optical R-band \citep{2018A&A...615A.118S,2020ApJ...895..122C}. Recently \citet{2021A&A...645A.137A} reported a median period of 2.21$\pm$0.04 years using the historical optical light curves which  confirms the results of \citep{2015ApJ...813L..41A}; an additional secondary period of about 210 days was also detected and a spectral index $\alpha$ = 0.89$\pm$0.06 was found \citep{2021A&A...645A.137A}. \\ 
\\
This paper is organised as follows: Section 2 provides an overview of the telescopes, photometric observations, and the data reduction procedure. Analysis techniques we used to search for flux variability and correlations between bands are discussed in Section 3. Results of our study are reported in Section 4. A discussion and conclusions  are provided in Section 5.

\begin{table*}
\centering
\caption{Details of telescopes and instruments used} 
\label{tab:tel_log}             
\begin{tabular}{l c c c c c}          
\hline                		
             				& A1           	 & A2		    & AS1			& AS2		  	   & G						\\  \hline                           
Telescope  	& 1.30 m DFOT   	 & 1.04 m ST		& 60cm ASV     	&   1.4m ASV       & 70 cm meniscus telescope	\\         
CCD Model   & Andor 2K CCD     & STA4150      	    & CCD FLI PL230  	&	Andor iKon-L	   & FLI-4240				\\
Chip Size (pixels) 	& 2048$\times$2048      & 4096$\times$4096    	&  2048$\times$2064    	&	2048$\times$2048	   & 2048$\times$2048			\\
Scale (arcsec/pixel)   	    & 0.535  	     & 0.264   		&	0.518	    &	0.244		   &	 2.4						\\ 
Field (arcmin$^2$)    	    & 18$\times$18 		     & 16$\times$16        	&  17.7$\times$17.8    	&	8.3$\times$8.3	  	   &	 15$\times$15					\\ 
Gain (e$^-$/ADU)	    			& 2.0   		     & 3.49    	    	&	 1.8	  	    &	1			   &	 8						\\ 
Read-out Noise (e$^-$ rms)  	& 7.0 		     & 6.98    	    	&	 18.6  	  	&	7			   &	 10						\\ 
Typical Seeing (arcsec)     	& 1.2-2.0 	     & 1.2-2.5      	&	 1-2	      	&	1-2			   &	 1.0-2.0					\\ 
\hline                		
             				&R1				&R2				& B			    & BS					   & S\\   		
\hline                           
Telescope   		        & 50/70cm NAO	& 2m RC NAO	   	& 60cm AO		&Baldone Schmidt	   & 1.3m Modified RC	\\         
CCD Model   		        & FLI PL16803	&VersArray:1300B	&FLI PL9000		& STX-16803		   & Andor CCD DZ936-BXDD \\
Chip Size (pixels$^2$) 	& 4096$\times$4096		&1340$\times$1300   &3056$\times$3056	& 4096$\times$4096	& 2048$\times$2048		    \\
Scale (arcsec/pixel)   	    & 1.079			&	0.258		& 0.33	      	& 0.78				   & 0.2829				\\ 
Field (arcmin$^2$)    	    & 73.66$\times$73.66 &	5.76$\times$5.76 &16.8$\times$16.8 & 53$\times$53 & 9.6$\times$9.6				\\ 
Gain (e$^-$/ADU)	    			& 1				&	1		    & 1      		& 1.33				   & 0.8650				\\ 
Read-out Noise (e$^-$ rms)  	& 9				&	2			& 9         		& 10				   & 5.95				\\ 
Typical Seeing (arcsec)     	& 2-4			&	1.5-2.5		& 1.5-2.5     	& 2-4				   & 1-1.3					\\ 
\hline          
\end{tabular}

\footnotesize
A1: 1.3-m Devasthal Fast Optical Telescope (DFOT) at ARIES, Nainital, India\\
A2: 1.04-m Samprnanand Telescope(ST), ARIES, Nainital, India\\
AS1: 60-cm Cassegrain Telescope, Astronomical Station Vidojevica (ASV), Serbia \\
AS2: 1.4-m telescope located at Astronomical Station Vidojevica, Serbia\\
G: 70-cm meniscus telescope at Abastumani Observatory, Georgia  \\
R1: 50/70-cm Schmidt Telescope at  National Astronomical Observatory Observatory, Rozhen, Bulgaria   \\
R2: 2-m Ritchey-Chretien telescope at National Astronomical observatory Rozhen, Bulgaria\\
B: 60-cm Cassegrain Telescope at Astronomical Observatory Belogradchik, Bulgaria \\
BS: 1.20-m Baldone Schmidt Telescope at the Institute of Astronomy, University of Latvia\\ 
S: 1.3-m Skinakas Observatory, Crete, Greece \\
\end{table*}

\section{Observations and Data Reduction} \label{sec:data}
\noindent
In 2019, the TeV blazar PG 1553$+$113 was observed using 10 optical ground-based telescopes located in 6 countries (India, Georgia, Greece, Bulgaria, Serbia, and Latvia). Key details about these telescopes are given in Table \ref{tab:tel_log}. These telescopes are equipped with CCD detectors and UBVRI broadband optical filters \citep{2012MNRAS.420.3147G,2015MNRAS.452.4263G,2016MNRAS.458.1127G}. Extensive photometric observations of PG 1553$+$113 were carried out using these 10 telescopes  from late February to early November 2019. The detailed observation log is given in Table \ref{tab:obs_log}.\\ \\
The data obtained from the Indian, Latvian and Greek telescopes were processed using the same data reduction steps. For the processing of the raw data, we used standard procedures of the Image Reduction and Analysis Facility (IRAF)\footnote{IRAF is distributed by the National Optical Astronomy Observatories, which are operated by the Association of Universities for Research in Astronomy, Inc., under cooperative agreement with the National Science Foundation.} software following the steps described below. For image pre-processing, we generated a master bias frame for each observing night which was subtracted from all twilight flat frames and all source image frames taken on that particular night. A master flat was generated for each filter by taking the median of all the bias subtracted twilight sky flat frames and then normalising the master flat. To remove pixel-to-pixel inhomogeneity, the source image was divided by the normalised master flat of the same filter. Finally, cosmic ray removal was carried out for all source image frames. To find the instrumental magnitudes of the blazar PG 1553$+$113 and its local standard stars \citep[see Figure 1,][]{2015MNRAS.454..353R} we employed a concentric circular multi-aperture photometry technique using the DAOPHOT software II\footnote{Dominion Astrophysical Observatory Photometry software} \citep{1987PASP...99..191S,1992ASPC...25..297S}.  Explicitly, for the aperture photometry, we took four different concentric aperture radii i.e., 1 $\times$ FWHM (Full Width at Half Maximum), 2 $\times$ FWHM, 3 $\times$ FWHM, and 4 $\times$ FWHM for every night. In several earlier studies, we have found that the aperture radii = 2 $\times$ FWHM provides the best S/N \citep[e.g.,][references therein]{2016MNRAS.458.1127G,2019ApJ...871..192P}, so we used this value for our final analysis. Image processing of data obtained from the Bulgarian and Georgian  telescopes was performed using ESO-MIDAS2\footnote{ESO-MIDAS is the European Southern Observatory Munich Image Data Analysis System which is maintain and  developed by European Southern Observatory}\citep{2012MNRAS.420.3147G} and MAXIMDL\footnote{\url{https://diffractionlimited.com/help/maximdl/MaxIm-DL.htm}} is used for the Serbian telescopes \citep{2020MNRAS.496.1430P}, respectively, in a similar fashion that IRAF was used on the observations from the Indian, Latvian and Greek telescopes. \\
\\
In every night of observations we observed all 4 local standard stars on the same field  \citep[see Figure 1 of][]{2015MNRAS.454..353R}. We selected two standard stars, Stars 2 and 3 from Figure 1 of \citep{2015MNRAS.454..353R}). These two standard stars have magnitudes and colours close to that of the target blazar. Finding such ideal standard stars in blazar fields are rare. Using these stars for calibrating the blazar magnitude will avoid any error occurring from differences in the photon statistics in the differential photometry of the blazar and standard stars. Since the magnitudes of PG 1553$+$113 and the standard stars were obtained simultaneously under the same air mass and weather conditions, there is no need for correction of atmospheric extinction. Finally one comparison star (Star 2) was used to calibrate the instrumental magnitude of PG 1553$+$113. 

\section{ANALYSIS TECHNIQUES} \label{sec:AnaTech}
\noindent
To examine the frequency of intra-day variability of the blazar PG 1553$+$113 from these optical data, we have employed two different analysis techniques: power enhanced F-test and nested analysis of variance (ANOVA). These are more reliable methods for quantifying variabilty than the earlier widely used statistical tests such as the standard F-test and C-test \citep{2014AJ....148...93D,2015AJ....150...44D}. Both of these methods use several comparison stars in the analysis. We also examine the amplitudes and duty cycles of its optical variability and perform cross-correlation analyses between bands.

\subsection{Power-enhanced F-test}
\label{sec:f_test}
To obtain the optical intra-day variability in the blazar PG 1553$+$113, our use of the power-enhanced F-test followed the approach given in \citet{2014AJ....148...93D} and \citet{2015AJ....150...44D}. This test has frequently been used in recent studies for finding IDV in blazars \citep[e.g.,][and references therein]{2015MNRAS.452.4263G,2016MNRAS.460.3950P,2017MNRAS.466.2679K,2019ApJ...871..192P,2020MNRAS.496.1430P}. In this test we compare the variance of the source LC to the combined variance of those of all standard stars. We used the closest brightest star to the object as our reference star to minimize errors. The other standard stars in the blazar field are considered as the comparison stars. The enhanced F-test is defined as 
$F_{enh} = {s_{blz}^2}/{s_c^2}$ \citep{2014AJ....148...93D} 
where $s_c^2 = \frac{1}{(\sum_{j=1}^k N_j)-k}\sum_{j=1}^{k} \sum_{i=1}^{N_i} s_{j,i}^2$. 
Here, s$_{blz}^{2}$ is the variance of  differential light curve (DLC) of the  blazar and reference star, and s$_{c}^{2}$ is the stacked variance of the comparison star-reference star DLCs \citep{2014AJ....148...93D}, where $N_j$ is the number of observations of the $j^{th}$ star and $k$ is the total number of comparison stars; s$_{j,i}^{2}$ is the scaled square deviation, which for the $j^{th}$ comparison star is defined as
$s_{j,i}^2 = \omega_j(m_{j,i}-\bar{m_j})^2$,
where $\omega_j$, $\bar{m_j}$, and m$_{j,i}$ , are the scaling factor, the mean magnitude of the jth comparison star DLC, and differential magnitude, respectively. The scaling factor $\omega_j$ is taken as the ratio of averaged square error of the blazar-reference star DLC to the averaged square error of the differential instrumental magnitudes in the comparison star-reference star DLC \citep{2011MNRAS.412.2717J}.
As $\omega_j$ is the ratio of the errors, it cancels out the too small (by factors of $\sim$1.5) photometric errors that emerge from IRAF  \citep{2004MNRAS.350..175S,2013JApA...34..273G}. The degree of freedom in the denominator is increased by the stacking of the variances of the comparison stars so the power of the power-enhanced F-test is enhanced when compared to the often  previously used simple F-test.\\
\\
In this work, we have three comparison field stars (S2, S3, and S4) from which star S2, having magnitude closest to the source's instrumental magnitude, is considered to be the reference star, and the remaining ($k=2$) field stars as the comparison stars. Since the blazar and all the comparison stars have the same number of observations ($N$), the number of degrees of freedom in the denominator and numerator in the power-enhanced F-statistics are $\nu_1 = N-1$ and $\nu_2 = k(N-1)$, respectively.
Then the $F_{enh}$ value is computed and compared with the critical value ($F_c$) at $\alpha = 0.01$, where $\alpha$ is the significance level set for the test, corresponding to a confidence level of $99\%$. Such a low  $\alpha$ value, indicates that the probability that the result arises by chance is very small. If $F_{enh}$ is larger than the critical value, the null hypothesis (no variability) is discarded. The estimated values of $F_{enh}$ and $F_{c}$ are given in Tables \ref{tab:var_res} and \ref{appendix:B1}.

\subsection{Nested ANOVA Test}
\label{sec:anova}
The one-way analysis of variance (ANOVA) test for AGN variability was introduced by \citet{1998ApJ...501...69D}. The nested ANOVA test is an updated version of the ANOVA test which uses several stars as reference stars to generate different differential light curves (DLCs) of the blazar. In contrast to power-enhanced F-test, no distinct comparison star is needed for the nested ANOVA test, and all the comparison stars are used as reference stars, so the number of stars in the analysis is increased by one \citep{2015AJ....150...44D,2019ApJ...871..192P}. The nested ANOVA test compares the average of dispersions between the groups of observations. In our case, we have used three reference stars (2, 3, and 4 which we call here as S2, S3, and S4, respectively)\footnote{\url{https://www.lsw.uni-heidelberg.de/projects/extragalactic/charts/1553+113.html}} to generate DLCs of the blazar. These three DLCs are then divided into a number of groups with four points in each group. A disadvantage of this technique is that microvariations shorter than the lapse time within each group of observations cannot be detected by the nested ANOVA test. But such phenomena, known as spikes, have been seldom reported in the literature \citep[e.g.][]{1996MNRAS.281.1267S,1998ApJ...501...69D,2004MNRAS.350..175S}.\\
\\
To get the results, we used our own Python program for this test. Following Equation (4) of \citet{2015AJ....150...44D} we calculated the mean square due to groups ($MS_G$) and mean square due to nested observations in groups ($MS_{O(G)}$). The ratio $F = MS_{G}/MS_{O(G)}$ follows an $F$ distribution with ${a-1}$ and $a(b - 1)$ degrees of freedom, in the numerator and denominator, respectively, where $a$ is the number of groups in an observation and $b$ is the number of data points in each group. A light curve is considered as variable if the value of F-statistic $\geq$ F$_c$ at 99 per cent confidence level, otherwise, we call it non-variable. The results of both the statistical tests are given in Tables \ref{tab:var_res} and \ref{appendix:B1}, where an LC is conservatively labeled as variable (V) only if both the tests found significant variations in it, otherwise it is labeled as NV, though of course there may be weak intrinsic variability even in some of those cases.

\subsection{Intraday Variability Amplitude}
For each of the variable light curves, we calculated the flux variability amplitude (Amp), using the equation given by \citep{1996A&A...305...42H}.
\begin{equation}
\label{sec:Intra}
Amp = 100\times \sqrt{(A_{max}-A_{min})^2 - 2 \sigma^2},
\end{equation}
where $A_{max}$ and $A_{min}$ are the maximum and minimum magnitudes, respectively, in the calibrated light curves of the blazar, while $\sigma$ is the mean error. The amplitude of variability is also mentioned in the last column of Table \ref{tab:var_res} for the variable light curve.

\subsection{Duty Cycle}
\noindent
The duty cycle (DC)  provides a direct estimation of the fraction of time for which a source has shown variability. We have estimated the DC of PG 1553$+$113 by using the standard approach \citep{1999A&AS..135..477R}. For these DC calculations, we considered only those LCs which were continuously monitored for at least about 1.5 hours, with
\begin{equation} 
DC = 100\frac{\sum_\mathbf{i=1}^\mathbf{n} N_i(1/\Delta t_i)}{\sum_\mathbf{i=1}^\mathbf{n}(1/\Delta t_i)}  {\rm per~cent} 
\end{equation}
here $\Delta t_i = \Delta t_{i,obs}(1+z)^{-1}$ is the redshift corrected observing time of the source during the $i^{th}$ observation, and $N_i$ takes the value 1 if IDV is detected or 0 if not detected. Computation of the DC has been weighted by the observing time $\Delta t_i$ for the $i^{th}$ observation, as the observation time is different for each observation.  

\subsection{Discrete Correlation Function} 
\label{sec:DCF}
There may be time lags between the observed light curves in different optical bands. To examine this possibility, we used the Discrete Correlation Function (DCF) given by \citet{1988ApJ...333..646E} which is an useful analysis tool for unevenly sampled astronomical data. To get better estimations of errors, the technique was modified by \citep{1992ApJ...386..473H}. Using the method, firstly we calculate the unbinned correlation (UDCF) using the given time series by
\begin{equation}
 UDCF_{ij}=\frac{(x(i) - \bar{x})(y(j) - \bar{y})}{\sqrt{\sigma_x^2 \sigma_y^2}},
\end{equation}

\noindent
where $\bar{x}$ and $\bar{y}$  are the mean values of the two discrete data series $x(i)$ and $y(j)$, with standard deviations $\sigma _x^2$ and $\sigma_y^2$ and measurement errors e$_{x}$, e$_{y}$. After calculation of the UDCF, the correlation function is binned. Averaging the UDCF values ($N$ in number) for each time delay, $\tau$, to calculate the DCF,

\begin{equation}
DCF(\tau)=\frac{1}{N}\sum  UDCF_{ij} .
\end{equation}

\noindent
where $\tau$ is the center of the bin of size $\Delta\tau$. The error is found from the standard deviations of the number of bins used for determining the DCF and is given as:

\begin{equation}
\sigma_{DCF}(\tau) = \frac{\sqrt{\sum[UDCF_{ij} - DCF(\tau)]^2}}{N - 1} .
\end{equation}

\noindent
A positive DCF peak implies that the data from the two different EM bands are correlated, while two data sets are anti-correlated if DCF value ${<}$ 0, but no DCF peak or DCF = 0 implies that no correlation exists between the two different EM bands data. When correlating a data series with itself (i.e., $x = y$), we obtain an autocorrelation function (ACF) with an automatic peak at $\tau$ = 0, indicating the absence of any time lag. For an ACF, any other strong peak can indicate the presence of periodicity \citep{2017ApJ...841..123P}. 

\begin{table*}
\caption{Results of IDV analysis of PG 1553$+$113}            
\label{tab:var_res}                   
\centering 
\begin{tabular}{lcccccccccc} \hline \hline                		 
Observation date & Band & \multicolumn{3}{c}{{\it Power-enhanced  F-test}}  & \multicolumn{3}{c}{{\it Nested ANOVA}} &Status &   Amplitude\\
\cmidrule[0.03cm](r){3-5}\cmidrule[0.03cm](r){6-8} yyyy-mm-dd & &DoF($\nu_1$,$\nu_2$ ) & $F_{enh}$ & $F_c$  & DoF($\nu_1$,$\nu_2$ ) & $F$ & $F_c$& & $\%$& \\\hline
20190417  & R     & 47, 94 & 2.78 & 1.76    & 11, 36 & 9.44 & 2.79 &  V & 14.98  \\
20190418  & R     & 102,204& 2.93 & 1.48    & 19,80  & 4.19 & 2.14 &  V & 8.89  \\
20190515  & V     & 35, 70 & 6.53 & 1.93    & 6, 28 & 10.36 & 3.52 &  V & 11.82  \\
		  & R     & 35, 70 & 3.13 & 1.93    & 6, 28 &  8.57 & 3.52 &  V & 9.02  \\
		  & I     & 35, 70 & 3.21 & 1.93    & 6, 28 &  3.75 & 3.52 &  V & 8.19  \\
20190613  & R     &123, 246& 1.97 & 1.43    & 23,96 & 2.65 & 2.00 &  V &  3.91 \\\hline
\end{tabular}
\end{table*}

\setcounter{figure}{0}
\begin{subfigures}
\label{fig:idv1}
\begin{figure*}
\centering
\includegraphics[width=18cm, height=20cm]{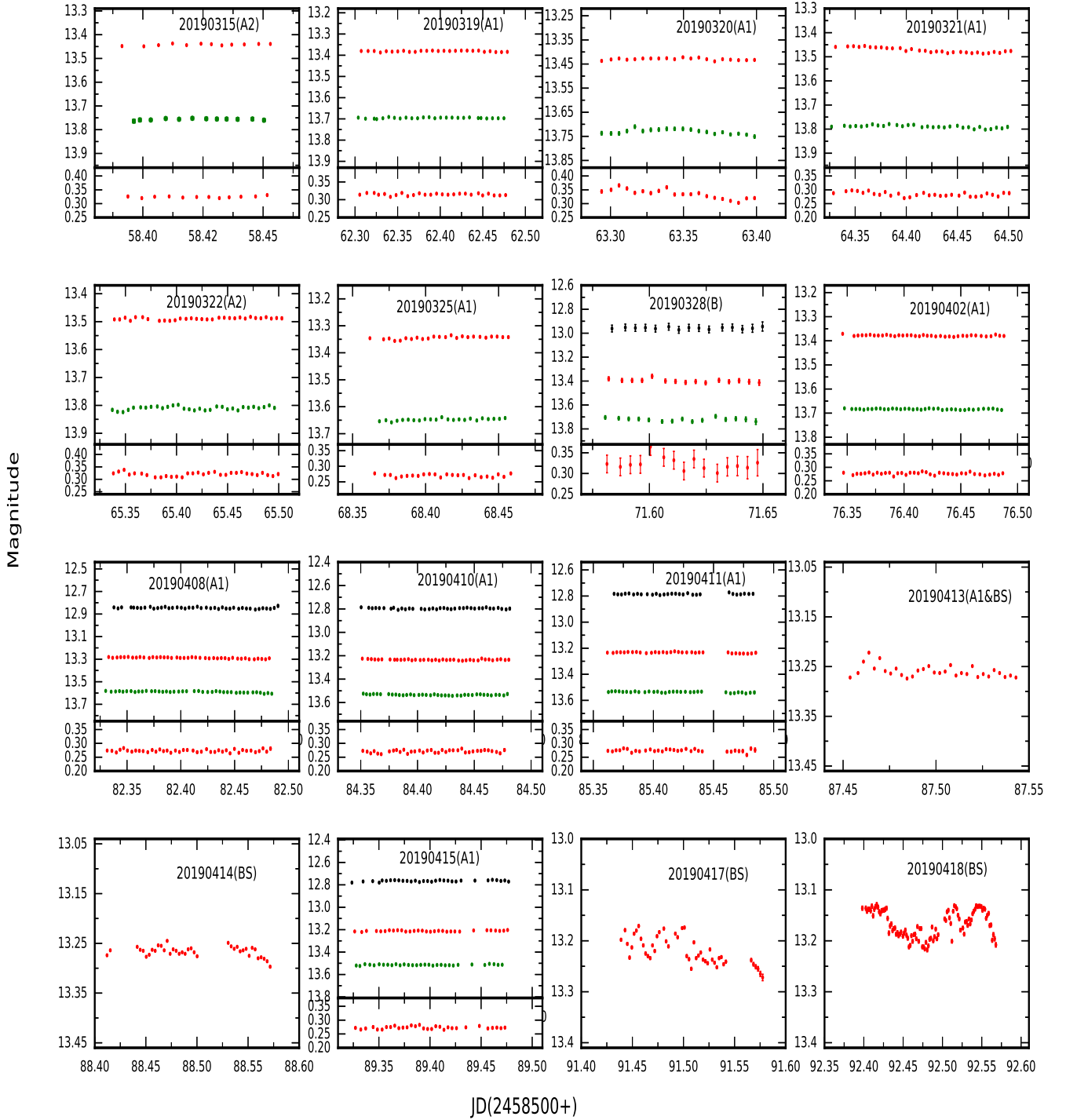}
\vspace*{-0.4in} 
\caption{\label{1_a} Upper sub-panels display IDV optical (VRI) light curves of PG 1553$+$113; they are shown in green  (V), red (R), and black (I), respectively;  bottom sub-panels show the colour (V$-$R) variation on IDV timescales. The date of observations and the telescope abbreviation are given in  each panel.} 
\end{figure*}

\begin{figure*}
\centering
\includegraphics[width=18cm, height=15cm]{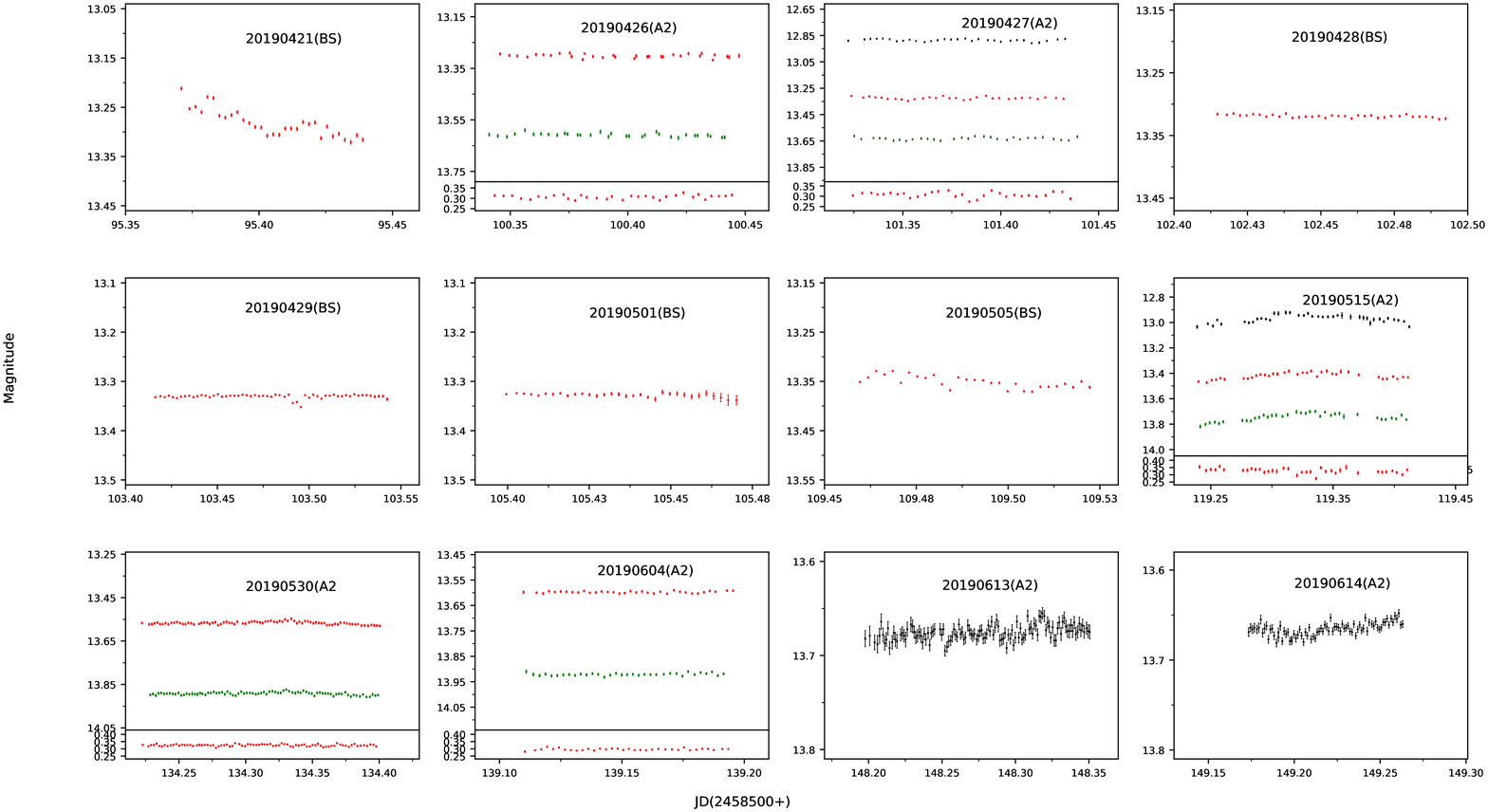}
\vspace*{-0.35in}
\caption{\label{1_b} Continued} 
\end{figure*}
\end{subfigures}

\section{Results}
\noindent
Our optical photometric observations of the TeV blazar PG 1553$+$113 were carried out during a total of 91 distinct nights from 28 February to 8 November 2019 using 10 optical telescopes listed in Table \ref{tab:tel_log}.
One or more of these telescopes observed the source quasi-simultaneously in B, V, R, and I bands in 43 nights, in V, R and I bands in 2 nights, in V and R bands in 1 night, and in only in the single R-band on 67 nights. In total, 2418 image frames were obtained during the complete observational period of which 77, 615, 1437, and 289 image frames were in the B, V, R, and I bands, respectively.  Now  we present the results of the variability properties of this blazar carried out on  IDV and STV timescales.

\subsection{Intraday Variability}
\subsubsection{Intraday Flux Variability}
\label{sec:flux}
\noindent
We plot the calibrated V, R, and I band magnitude versus time IDV LCs for the blazar PG 1553$+$113 in the upper panels of Figure \ref{fig:idv1}; when available, the V$-$R colour versus time plots are in the lower panels. 
To search for the clear presence of IDV, we performed the statistical tests discussed in sections \ref{sec:f_test} and \ref{sec:anova}. The complete results of this analysis are given in Table \ref{appendix:B1}.\\ 
 \\
 Significant IDV was detected in R-band LCs of PG 1553$+$113 on April 17, 18 and June 13. On May 15, we found IDV in all of the V, R, and I bands. Notice that the errors in the V-band LCs are roughly twice as large as those in R, so this reduces the likelihood of detecting any small variations that might be present. We also estimated the IDV amplitudes for the confirmed variable LCs, shown in the last column of Table \ref{tab:var_res} using Equation \ref{sec:Intra}. 
  Usually a blazar's variability amplitude is larger at higher frequencies, as seen in this case on 15 May 2019 in which I, R, and V band all showed variability. This trend suggests that the blazar spectrum gets steeper with decreasing brightness and flatter with increasing brightness \citep[e.g.,][]{1998MNRAS.299...47M,2015MNRAS.450..541A}. 
 However, on some occasions the variability amplitude of blazars at lower frequencies was found to be comparable or even larger than that at higher frequencies \citep[e.g.,][]{2000ApJ...537..638G,2015MNRAS.452.4263G}. 
 
\begin{figure*}
\centering
\vspace*{-4.0in}
\includegraphics[width=18cm, height=18cm]{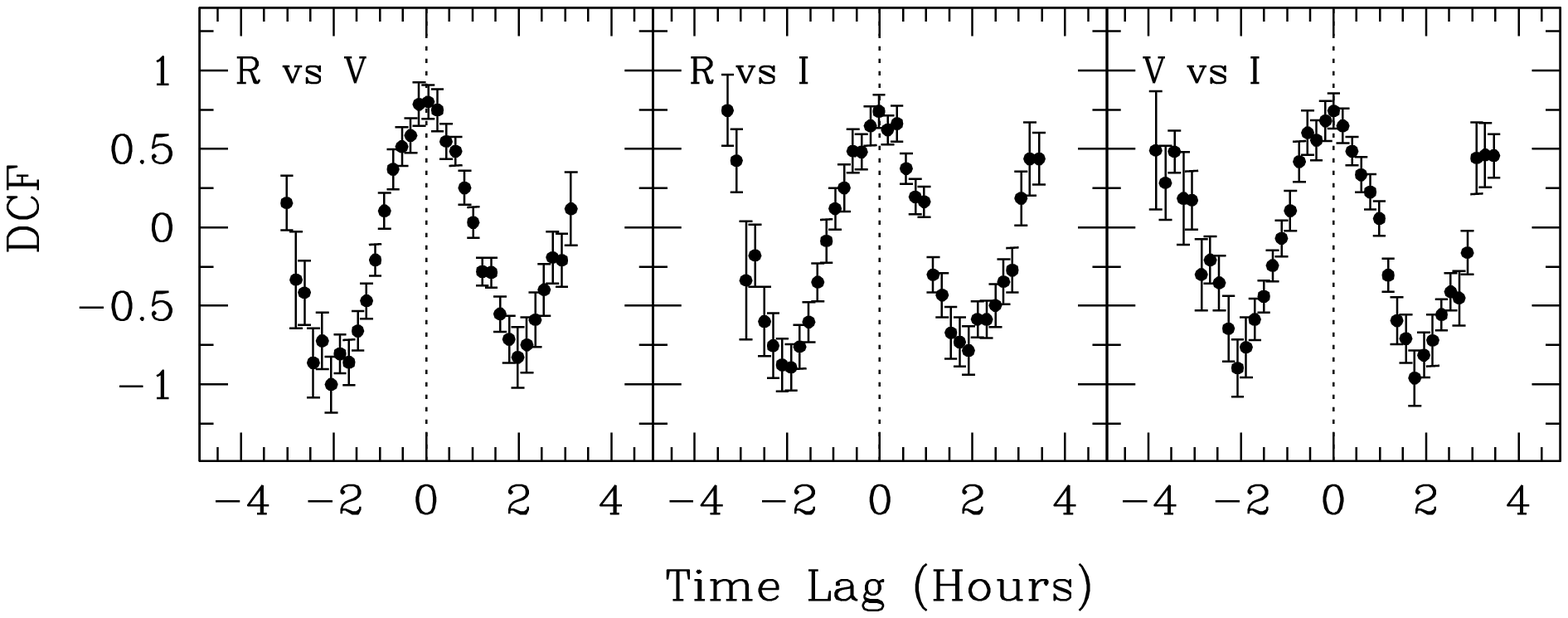}
\vspace*{-0.5in}
\caption{DCF plots for PG 1553$+$113 on 2019 May 15 showing peaks consistent with 0 lag. In each plot, X- and Y-axis are the time lags in hours and DCF values, respectively.}
\label{fig:dcf}
\end{figure*}
 
 \subsubsection{Intraday Cross-correlated Flux Variability}
 \noindent
We have used the DCF technique as described in Section \ref{sec:DCF} to determine the cross correlations and thus search for any time lags between the V, R, and I optical bands for  2019 May 15, the only night in which we observed IDV variation from all bands. We took a DCF bin size of 12 minutes and get strong peaks (with DCFs $\sim$ 1) at an essentially  zero lag, as is clear from  Figure \ref{fig:dcf}. The null time lags imply that the photons in these wavebands are emitted by the same physical processes and from the same emitting region, which is not surprising, given how close in frequency these optical bands are used the interpolated cross-correlation function and found no significant time lag between the different optical bands \citep[e.g.,][and references therein]{2015MNRAS.450..541A,2016MNRAS.455..680A,2017MNRAS.471.2216B}. 

\subsubsection{Intraday Duty Cycle} 
\label{sec:DC}
\noindent
\citet{2021A&A...645A.137A} carried out an optical IDV literature survey which included 28 new optical IDV light curves of PG 1553$+$113. A total of 74 optical IDV observations carried out during 1999--2019 were discussed there and  in them the source showed IDV in 8 nights i.e., DC = 10.8\%. In our current work we have 50 IDV LCs taken in 27 observing nights. We found 4 variable IDV LCs out of 27 LCs in the R-band, 1 variable IDV LC out of 17 LCs in the V-band, and 1 variable IDV LC out of 6 LCs in the I-band. The DCs for the R, V, and I bands are 14.8\%, 5.88\%, and 16.7\%, respectively. Combining them, the DC of PG 1553$+$113 based on all these new IDV LCs is 12\%, which is consistent with the previous results \citep[e.g.,][and references therein]{2021A&A...645A.137A}.    

\begin{figure*}
\centering
\includegraphics[width=18cm, height=20cm]{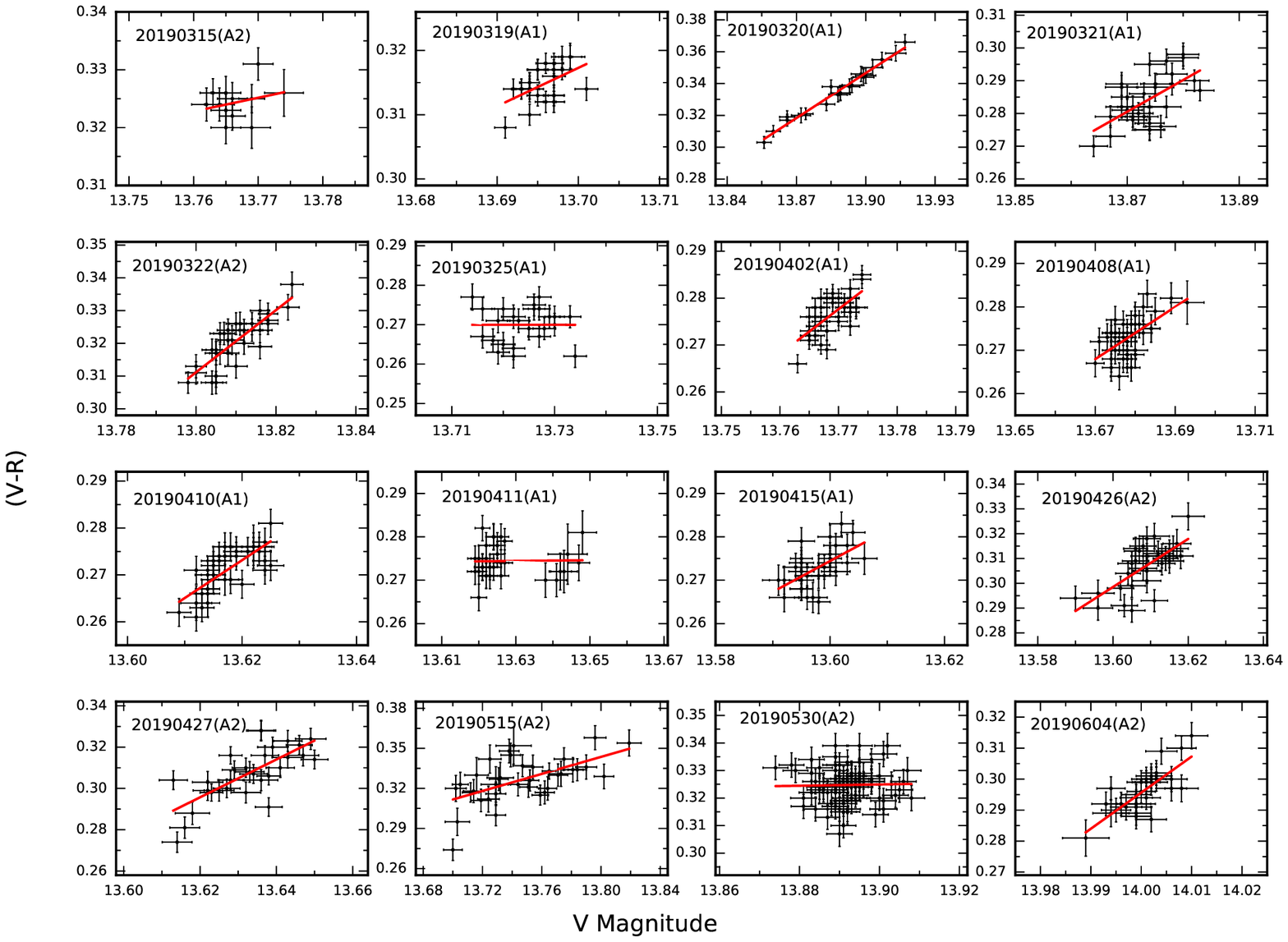}
\vspace*{-0.4in}
\caption{IDV colour$-$magnitude plots for PG 1553$+$13. The observation date and the telescope code are given in each plot.}
\label{fig:col_mag}
\end{figure*}

\begin{table}
\caption{Linear fits to colour-magnitude plots}            
\label{tab:r_v}                   
\centering    
\begin{tabular}{ccrcc} \hline \hline                		 
Observation date & $m_1^a$ &  $c_1^a$ & $r_1^a$ &  $p_1^a$  \\
yyyy-mm-dd   &         &          &         &   \\\hline            
2019-03-15	&0.236 & -2.927   & 0.267  & 3.968e-01 \\
2019-03-19	&0.596 & -7.847   & 0.495  & 1.021e-02 \\
2019-03-20	&0.942 & -12.752  & 0.988  & 3.677e-16 \\
2019-03-21	&0.968 & -13.158  & 0.596  & 4.971e-04 \\
2019-03-22	&0.948 & -12.796  & 0.838  & 3.876e-09 \\
2019-03-25	&0.000 & 0.272    & 0.000  & 9.992e-01 \\
2019-04-02	&0.955 & -12.874  & 0.675  & 4.429e-06 \\
2019-04-08	&0.602 & -7.972   & 0.607  & 2.533e-05 \\
2019-04-10  &0.805 & -10.702  & 0.742  & 1.405e-07 \\
2019-04-11  &0.008 & 0.162    & 0.019  & 9.151e-01 \\
2019-04-15  &0.713 &-9.392    & 0.576  & 6.848e-04 \\
2019-04-26  &0.973 &-12.895   & 0.699  & 8.376e-06 \\
2019-04-27  &0.915 &-12.172   & 0.723  & 6.291e-06 \\
2019-05-15  &0.319 &-4.059    & 0.585  & 1.775e-04 \\
2019-05-30  &0.023 &-0.002    & 0.024  & 8.296e-01 \\
2019-06-04  &1.164 &-16.009   & 0.765  & 1.301e-07 \\\hline                           
\end{tabular}\\
$^am_1$ = slope and $c_1$ = intercept of CI against V-mag; $r_1$ = Correlation coefficient; $p_1$ = null hypothesis probability
\end{table}

\subsubsection{Intraday Colour Variation} 
\label{sec:ICV}
\noindent
To study colour variation of the TeV blazar PG 1553$+$113 on IDV timescales  with respect to time and V-band magnitude (colour magnitude variation), we can use the 16 nights of data on which quasi-simultaneous observations were carried out in V and R-bands. We calculated the V$-$R colour indices (CIs)  for each pair of V and R magnitudes and  plotted these V$-$R CIs with respect to time in the bottom panel of Figure \ref{fig:idv1}, and our results are presented in the last column of Tables \ref{tab:var_res} and  \ref{appendix:B1}.\\ 
\\
To investigate the colour behaviour of PG 1553$+$113 with respect to V-band magnitude (colour-magnitude (CM) plot), we fitted a straight line of the form CI = mV + C on each CM plot, which are displayed for each night in Figure \ref{fig:col_mag}, with the results listed in Table \ref{tab:r_v}. Clear global brighter-when-bluer (BWB) trends were observed in 11 nights on IDV timescales, with correlation coefficients ranging between 0.5 to 0.9. Hence this source's brightness was found to have strong BWB chromatism on IDV timescales. Such a BWB trend has been found to be predominant in BL Lac objects in the large number of optical observations made during both flaring and steady states \citep[e.g.,][]{2000ApJ...537..638G,2006A&A...450...39G,2012MNRAS.425.3002G}. \citet{2004A&A...421..103V} found that the intra-day flares followed a BWB trend strongly, with a slope of $\sim$ 0.4, which is also the case for our data, with the slopes of the IDV LCs more than $\sim$ 0.5 in majority of cases.\\
\\
A bluer-when-brighter (BWB) trend is commonly observed in blazars \citep[e.g.,][]{2001A&A...377..396R,2002A&A...390..407V,2003A&A...397..565P,2001AJ....121...90C,2007A&A...470..857P,2010MNRAS.404.1992R,2015MNRAS.450..541A}, in particular blazars of the HSP class, in which the optical continuum is generally believed to be entirely dominated by the non-thermal jet synchrotron emission, and can be explained in shock-in-jet models. In LSP blazars, in which the accretion disc can provide a contribution to the optical continuum, a redder-when-brighter (RWB) trend sometimes is seen and probably indicates a relatively increasing thermal contribution at the blue end of the spectrum when the  non-thermal jet emission decreases
\citep[e.g.,][]{2006A&A...453..817V,2007A&A...464..871R,2012MNRAS.425.3002G}. 

\begin{figure*}
\centering
\includegraphics[width=18cm, height=10cm]{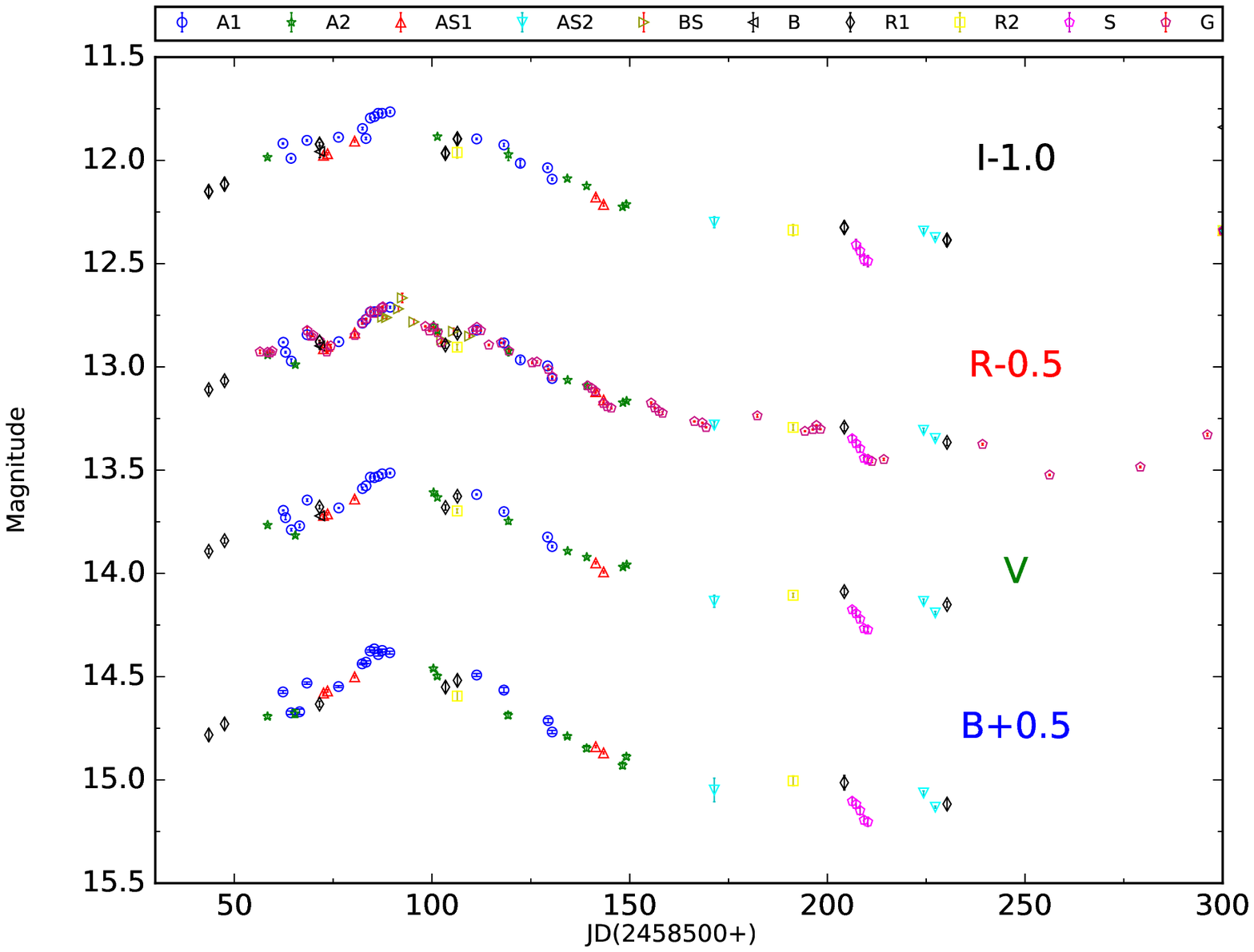}
\vspace*{-0.35in}
\caption{STV optical (BVRI) light curves of PG 1553$+$113, the telescopes, identified by different colours and symbols, are noted at the top of the figure.}
\label{fig:stv}
\end{figure*}
 
\subsection{Short Term Variability}
\noindent
\subsubsection{Short Term Flux Variability}
\label{sec:STFV}
\noindent
 The STV light curves of PG 1553$+$113 in B, V, R, and I bands for the entire monitoring period are shown in Figure \ref{fig:stv}, where we have plotted the nightly averaged magnitudes in B, V, R, and I bands with respect to time. During our monitoring period the source was detected in the brightest state of R = 13.13 mag on April 18, while the faintest level detected was R = 14.02 mag on September 29. The mean magnitudes were 14.23, 13.83, 13.51 and 13.07 in B, V, R, and I bands, respectively. The variability on STV timescales can be clearly seen at all optical wavelengths. Using equation \ref{sec:Intra}, we have estimated very similar variability amplitudes of 83.7 \%, 75.9 \%, 75.8 \%, and 72.3 \%, respectively, in B, V, R, and I bands.

\begin{figure}
\centering
\includegraphics[width=8cm, height=8cm]{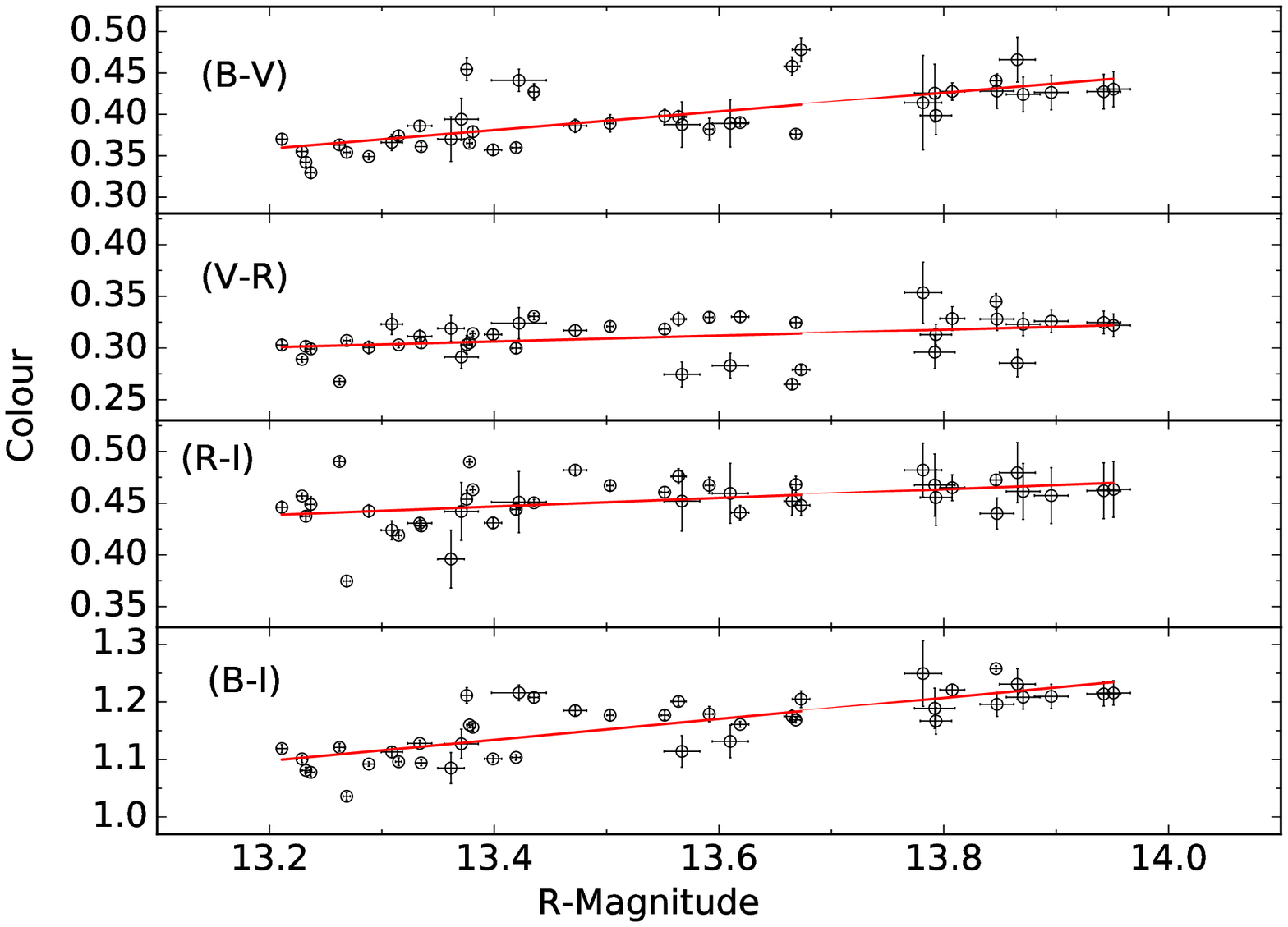}
\vspace*{-0.1in}
\caption{Optical colour–magnitude plots of PG 1553$+$113 during our total observation duration.}
\label{fig:ST_SI1}
\end{figure}

\begin{table}
\caption{Colour–Magnitude Dependencies and Colour–Magnitude Correlation Coefficients on Short Timescales}
\label{tab:Var_CM}                   
\centering    
\resizebox{0.35\textwidth} {!}{  
\begin{tabular}{ccccc} \hline \hline                		 
Colour Index & $m_1^a$ &  $c_1^a$ & $r_1^a$ &  $p_1^a$  \\\hline 
B$-$I	&0.182 		 &-1.308 		&0.775   		&1.7e-07		\\ 
R$-$I &0.041 		 &-0.109 	    &0.418   		&0.006		\\	 
V$-$R	&0.028 	     &-0.074 		&0.332 			&0.032   \\		 
B$-$V	&0.112     	 &-1.125 	    &0.714   		&1.1e-07		\\
\hline                           
\end{tabular}}\\
$^am_1$ = slope and $c_1$ = intercept of CI against R-mag; $r_1$ = Correlation coefficient; $p_1$ = null hypothesis probability
\end{table}

\begin{figure}
\centering
\includegraphics[width=8cm, height=8cm]{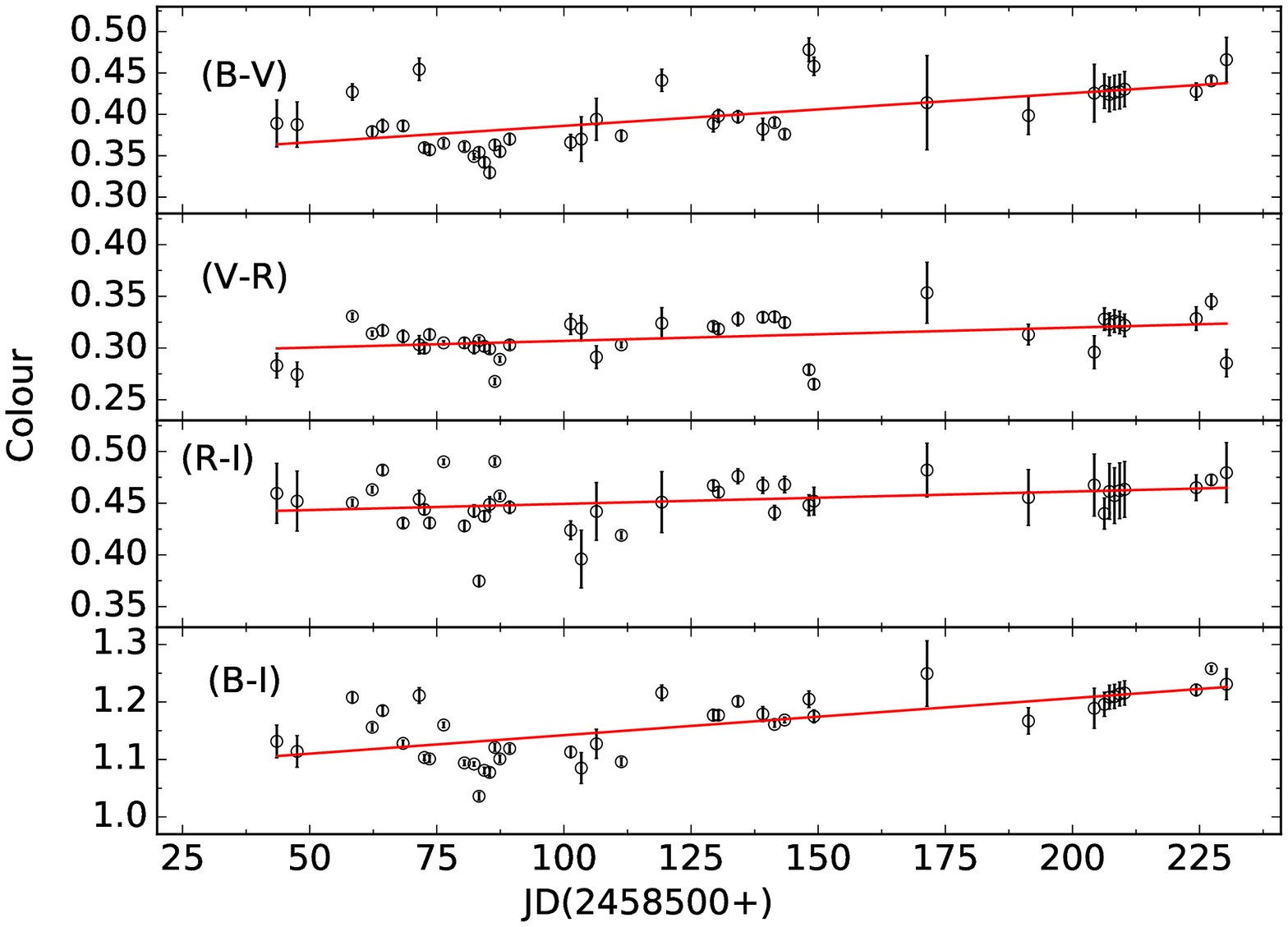}
\vspace*{-0.1in}
\caption{Optical colour variability LCs covering the total observation duration of PG 1553$+$113}
\label{fig:ST_SI2}
\end{figure}

\begin{table}
\caption{Colour Variation with Respect to Time on STV Timescales}
\label{tab:Var_JD}                   
\centering    
\resizebox{0.35\textwidth} {!}{  
\begin{tabular}{ccccc} \hline \hline                		 
Colour Index & $m_1^a$ &  $c_1^a$ & $r_1^a$ &  $p_1^a$  \\\hline 

B$-$I	&	0.0006 	 &1.077 			&0.678 			&7.9e-05		\\
R$-$I	&	0.0001 	 &0.437 			&0.297 			&0.055			\\
V$-$R	&	0.0001 	 &0.294 			&0.374			&0.015			\\
B$-$V   &	0.0004	 &0.346 			&0.622			&1.2e-05		\\ \hline
\end{tabular}}\\
$^am_1$ = slope and $c_1$ = intercept of CI against Time; $r_1$ = Correlation coefficient; $p_1$ = null hypothesis probability
\end{table}

\subsubsection{Spectral Variability}
\label{sec:SV}
\noindent
The colour-magnitude (CM) relationship can be useful for exploring various variability scenarios and better understanding the origin of blazar emission. Therefore, we searched for any relationship of the source's colour indices (CIs) with brightness in the R-band and with respect to time. We fitted the plots of the optical CIs such as (B$-$V), (B$-$I), (V$-$R) and (R$-$I) with respect to both R-band magnitude and time along a straight line of the form Y = mX + C as shown in Figure \ref{fig:ST_SI1} and \ref{fig:ST_SI2}, respectively \citep{2020ApJ...890...72P}. Values of the parameters associated with these colour-time and colour-magnitude plots are given in Tables \ref{tab:Var_CM} and \ref{tab:Var_JD}, respectively. We found B$-$V and B$-$I colours to show highly significant variation with time as well as R magnitude, while the CIs involving the R-band show weaker trends in the same directions. A positive slope defines a significant positive correlation between CIs and blazar R magnitude, meaning that the source follows a bluer-when-brighter (BWB) trend, while a negative slope defines redder-when-brighter (RWB) trend \citep[e.g.,][and references therein]{2017MNRAS.465.4423G}. 

\begin{subfigures}
\label{fig:sed1}

\begin{figure}
\centering
\vspace*{-0.15in}
\includegraphics[width=9.5cm, height=9cm]{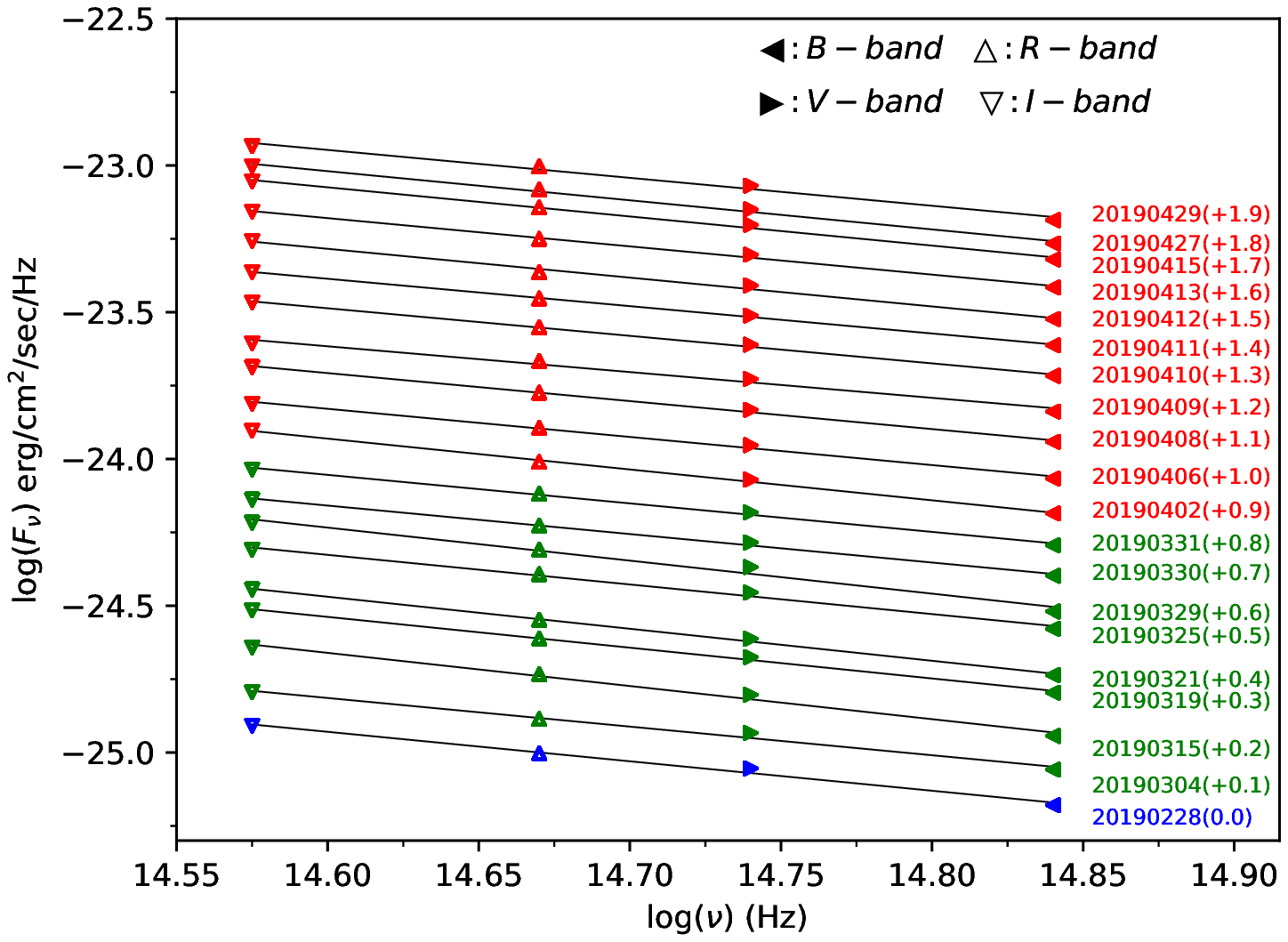}
\vspace*{-0.3in}
\caption{\label{8_a}The SED of PG 1553$+$113 in B, V, R, and I bands.}
\end{figure}

\begin{figure}
\centering
\vspace*{-0.15in}
\includegraphics[width=9.5cm, height=9cm]{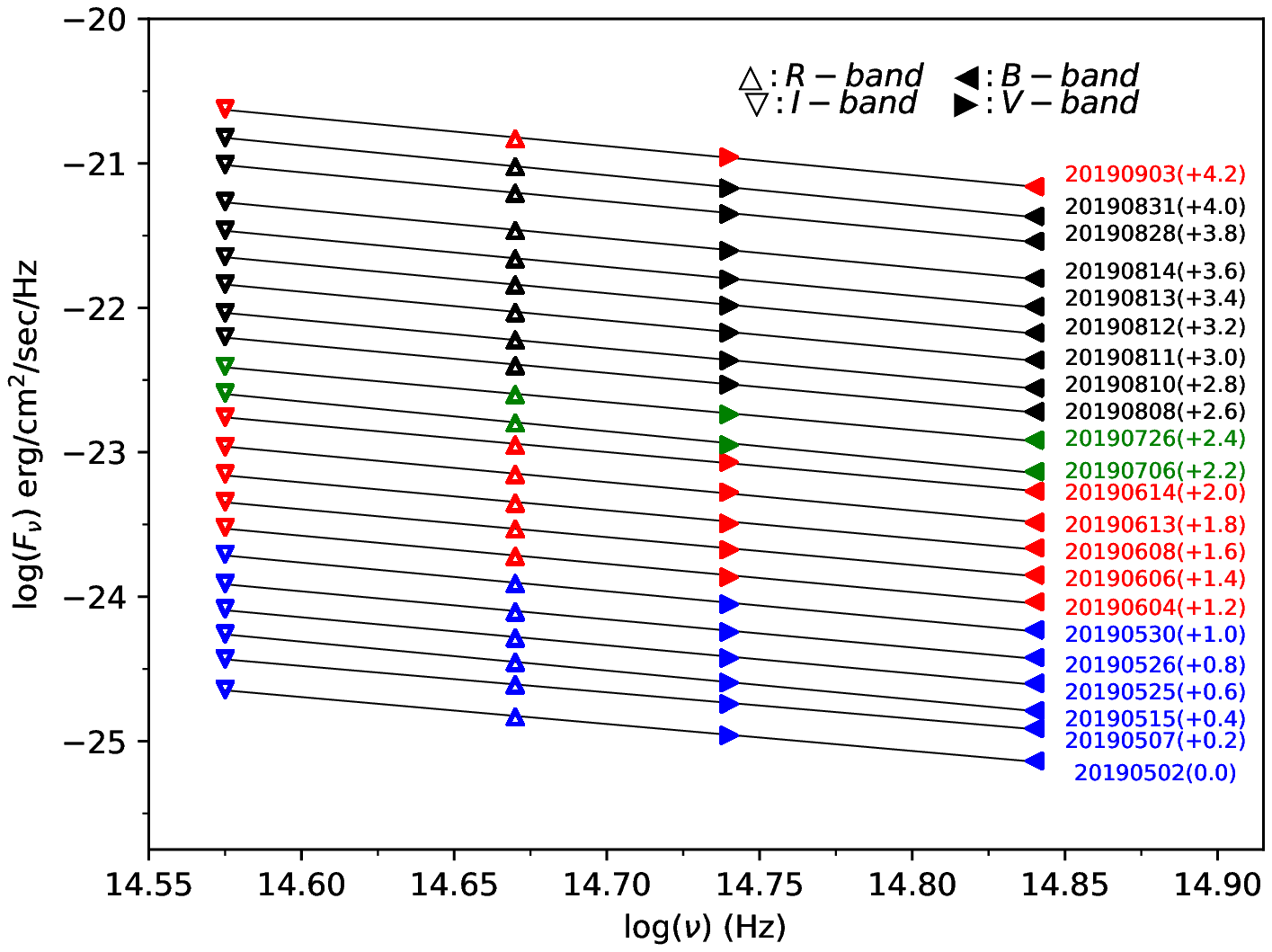}
\vspace*{-0.3in}
\caption{\label{8_b} Continued}
\end{figure}
\end{subfigures}

\begin{table*}
\caption{Straight-line Fits to Optical SEDs of TeV Blazar PG 1553$+$113}  
\label{tab:r_v1}                   
\centering    
\hskip-1.cm              
\begin{tabular}{ccrccccrcc}           
\hline                		 
Observation date & $m_1^a$ &  $c_1^a$ & $r_1^a$ &  $p_1^a$  & Observation date & $m_1^a$ &  $c_1^a$ & $r_1^a$ &  $p_1^a$  \\
yyyy-mm-dd   &         &          &         &               & yyyy-mm-dd   &         &          &         &               \\	\hline \hline 
2019-02-28	&-1.004 	& -10.276   & -0.995  & 0.004		&	2019-05-07 	&-0.961	& -10.591   & -0.994  & 0.005 \\ 
2019-03-04	&-0.975	& -10.679   & -0.994  & 0.005		&	2019-05-15 	&-1.138	&  -7.835   & -0.993  & 0.007 \\                    
2019-03-15	&-1.129 	&  -8.371   & -0.994  & 0.005		&	2019-05-25 	&-1.081	&  -8.499   & -0.998  & 0.002 \\
2019-03-19	&-1.124 	&  -8.409   & -0.991  & 0.009		&	2019-05-26 	&-1.081 	&  -8.327   & -0.997  & 0.003 \\
2019-03-21	&-1.048 	&  -9.524   & -0.998  & 0.001 		&	2019-05-30 	&-1.117 	&  -7.589   & -0.998  & 0.002 \\
2019-03-25	&-0.958 	& -10.837   & -0.996  & 0.003	 	&	2019-06-04 	&-1.087 	&  -7.846   & -0.998  & 0.001 \\
2019-03-29 	&-1.091 	&  -8.944   & -0.998  & 0.001		&	2019-06-06 	&-1.062 	&  -8.022   & -0.996  & 0.003 \\
2019-03-30 	&-1.009 	& -10.092   & -0.995  & 0.004	 	&	2019-06-08 	&-1.071 	&  -7.711   & -0.998  & 0.001 \\
2019-03-31 	&-0.972 	& -10.692   & -0.997  & 0.002	 	&	2019-06-13 	&-1.106 	&  -6.998   & -0.984  & 0.016 \\
2019-04-02	&-1.051 	&  -9.491   & -0.998  & 0.001	 	&	2019-06-14 	&-1.058 	&  -7.493   & -0.985  & 0.015 \\
2019-04-06 	&-0.953 	& -10.884   & -0.998  & 0.001	 	&	2019-07-06 	&-1.196 	&  -5.321   & -0.998  & 0.002 \\
2019-04-08	&-0.938 	& -11.084   & -0.998  & 0.001	 	&	2019-07-26 	&-1.064 	&  -7.063   & -0.996  & 0.004 \\
2019-04-09 	&-0.931 	& -11.182   & -0.999  & 0.001	  	&	2019-08-08 	&-1.089 	&  -6.494   & -0.993  & 0.006 \\
2019-04-10 	&-0.992 	& -10.281   & -0.997  & 0.003	 	&	2019-08-10 	&-1.111 	&  -5.999   & -0.993  & 0.007 \\
2019-04-11 	&-0.951 	& -10.957   & -0.992  & 0.007	 	&	2019-08-11 	&-1.126 	&  -5.588   & -0.995  & 0.005 \\
2019-04-12 	&-0.981 	& -10.449   & -0.995  & 0.004	 	&	2019-08-12 	&-1.131 	&  -5.341   & -0.994  & 0.005 \\
2019-04-13 	&-0.992 	& -10.331   & -0.996  & 0.003	 	&	2019-08-13 	&-1.135 	&  -5.082   & -0.995  & 0.005 \\
2019-04-15 	&-0.962 	& -10.732   & -0.997  & 0.002 		&	2019-08-14 	&-1.137 	&  -4.860   & -0.995  & 0.005 \\
2019-04-27 	&-0.878 	& -11.994   & -0.991  & 0.009 		&	2019-08-28 	&-1.146 	&  -4.462   & -0.995  & 0.004 \\
2019-04-29 	&-0.971 	& -10.672   & -0.997  & 0.002 		&	2019-08-31 	&-1.153 	&  -3.404   & -0.995  & 0.004 \\
2019-05-02 	&-1.001	& -10.213   & -0.994  & 0.005	 	&	2019-09-03 	&-1.145 	&  -4.102   & -0.990  & 0.011 \\\hline                           
\end{tabular}\\
$^am_1$ = slope and $c_1$ = intercept of $log(F_{\nu})$ against $log(\nu)$; $r_1$ = Correlation coefficient; $p_1$ = null hypothesis probability
\end{table*}

\subsection{Spectral Energy Distribution (SED)} 
\label{sec:spec}

\begin{figure}
\centering
\includegraphics[width=9cm, height=9cm]{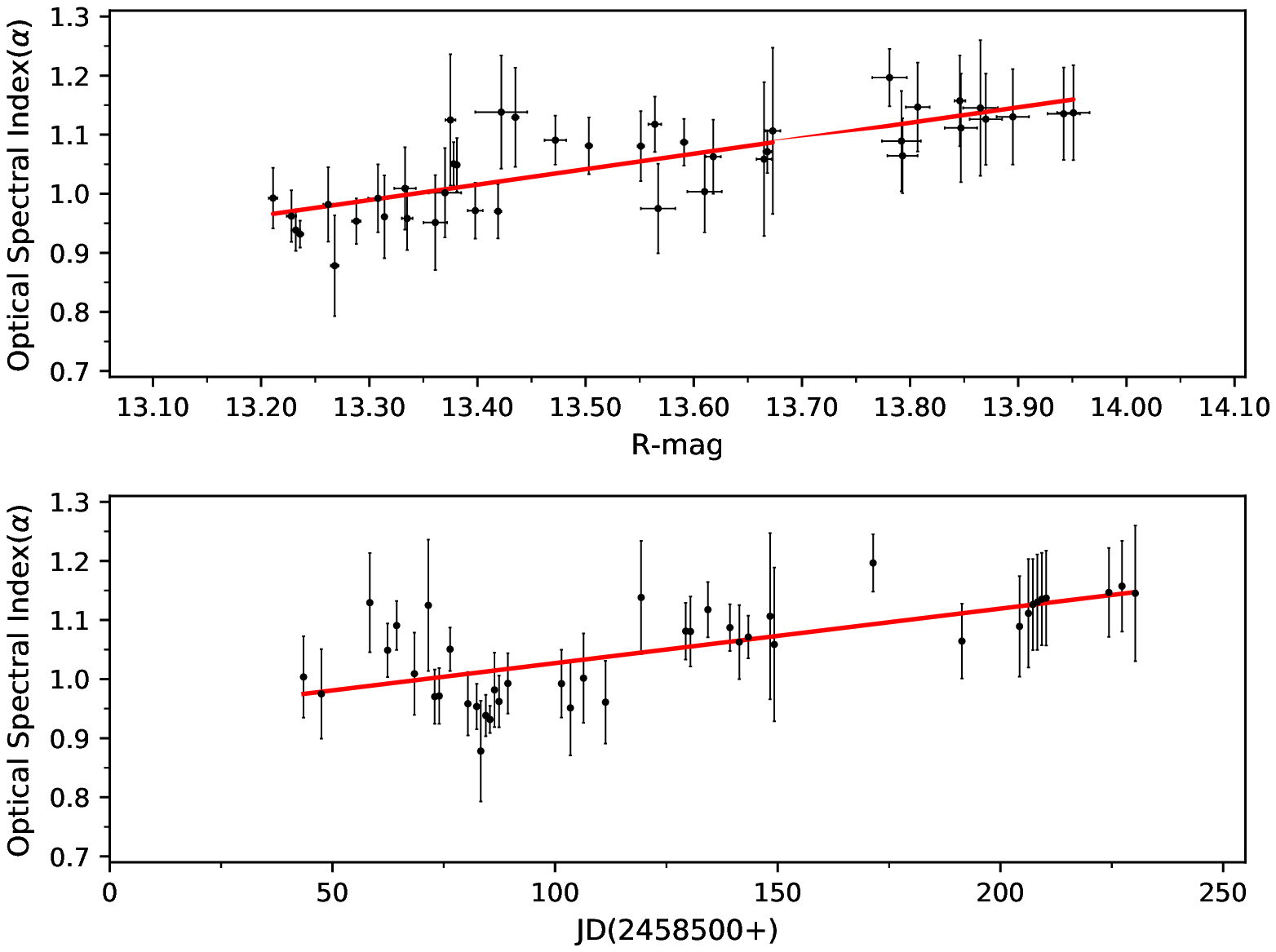}
\caption{Variation of optical spectral index ($\alpha$) of PG 1553$+$113 with respect to R-magnitude  (top) and time (bottom)}.
\label{fig:sed3}
\end{figure}

To study spectral variations during our observing period, we extracted the optical (BVRI) SEDs of the blazar for 42 nights in which observations were performed quasi-simultaneously in all four B, V, R, I wavebands. For this, we first dereddened the calibrated B, V, R, and I magnitudes by subtracting the Galactic extinction, with A$_{\lambda}$ having the following values: $A_B$ = 0.188 mag, $A_V$ = 0.142 mag, $A_R$ = 0.113 mag, and $A_I$ = 0.078 mag. The values of A$_{\lambda}$ were taken from the NASA Extragalactic Database  (NED \footnote{\url{https://ned.ipac.caltech.edu/}}). The dereddened calibrated magnitudes in each band were then converted into corresponding extinction corrected flux densities, $F_{\nu}$. We measured the source's brightest and faintest fluxes on 2019 Apr 18 and Aug 14, respectively. The optical SEDs of PG 1553$+$113, in log($\nu$) versus log($F_{\nu}$) representation, are plotted in Figure \ref{fig:sed1}. \\
\\
Since a simple power law ($F_{\nu} \propto \nu^{-\alpha}$), where $\alpha$ is known as the optical spectral index usually provides a good fit to the blazar optical continuum spectra \citep{2006MNRAS.371.1243H,2012MNRAS.425.3002G},
we fitted each SED with a first-order polynomial of the form log($F_{\nu}$) = $-\alpha$ log($\nu$) + C to get the optical spectral indices. The results of the fits are given in Table \ref{tab:r_v1} . \\
\\
The values of the spectral indices (${\alpha}$) range from 0.878 $\pm$ 0.029 to 1.106 $\pm$ 0.065 and their weighted mean was 0.94 $\pm$ 0.033. This value of the spectral index that we got is close to the results found earlier \citep{1994ApJS...93..125F,2021A&A...645A.137A}.  We show the spectral indices of PG 1553$+$113 with respect to time and R-band magnitude in the top and bottom panels of Figure \ref{fig:sed3}, respectively. We fitted each panel in Figure \ref{fig:sed3} with a first-order polynomial to investigate any variations in the spectral index. The optical spectral index increases with time and it also shows significant positive correlations with R-band magnitude. The study of MW SEDs can provide important information about physical parameters such as the magnetic field of the emitting region of the source.

\begin{table}
\caption {Variation of optical spectral index ($\alpha$), with respect to R-magnitude and STV timescale }
\label{tab:Alpha_T_JD}                   
\centering    
\resizebox{0.35\textwidth} {!}{  
\begin{tabular}{ccccc} \hline \hline                		 
Parameters              & $m$        &  $c$         & $r$           &$p$        \\\hline 
$\alpha$ vs Time   	&0.0009 	 &0.935		&0.673   		&1.04e-06    \\ 
$\alpha$ vs R$_{mag}$ &0.2608 	 &-2.492	    &0.771	    	&2.39e-08    \\\hline
\end{tabular}}\\
$m$ = slope and $c$ = intercept of $\alpha$ against R-mag and JD; $r$ = Correlation coefficient; $p$ = null hypothesis probability
\end{table}

\section{Discussion and Conclusions}
\noindent
In this work we analysed the optical photometric data from  PG 1553$+$113, a TeV blazar, collected using ten ground-based telescopes during February $-$ November 2019. In particular, we studied the flux and spectral variability properties of this blazar on both IDV and STV time-scales. We examined a total of 27 optical R-band, 17 in V-band and, 6 in I-band IDV LCs using two statistical methods: the power-enhanced F-test and the nested ANOVA test. We saw robust variability in 4 nights in R-band, and 1 in both V-band and I-band. Only one of these 27 intra-day LCs exhibits statistically significant IDV in all three bands and the amplitude of variability was only 11.82, 9.02, 8.19 \% respectively. 
So from our IDV analysis, we conclude that optical LCs of PG 1553$+$113 are either constant or show nominal variations on IDV time-scales.
As the blazar PG 1553$+$113 did not show large-amplitude variations during our monitoring period, we did not detect significant variations in colour with time (light curve) during our individual nightly observations.  There is one important caveat here:   since the durations of our observations are 1.5$-$5 hours, it is certainly possible that if we had longer nightly stares at this source we would have seen more frequent IDV.  For example, when the observation period was extended from $\leq$ 3 to $\geq$ 6 hours, the chances of IDV being found in a group of blazars increased from 64\% to 82\% \citep{2005A&A...440..855G}. We used a DCF analysis technique to search for any variability timescale or periodicity, as well as any possible correlation between different optical energy bands. We found a strong positive correlation, with no significant time lags between the different energy bands, which means that the emission region is the same these wavelengths. The correlation  between different energy bands also indicates that the emission is produced by the same electron populations. \\
\\
Flux variations on an STV timescale were also seen at all four optical wavelengths (I, R, V, and, B), while the colours were found to be non variable. An earlier optical photometric study of   PG 1553$+$113 was carried out by \citet{2012MNRAS.425.3002G}. They observed the blazar for IDV on six nights but found no significant IDV or colour (B$-$R) variability during any night. On STV timescales they detected genuine flux variability with no variation in colour. \citet{2016MNRAS.458.1127G} also monitored this source for IDV on 7 nights but found significant variations on IDV timescale only on one night. They reported significant flux variability with moderate colour variation on STV timescales. To find the optical spectral index($\alpha$) we made optical SEDs using quasi-simultaneous observation of B, V, R,and I optical band at different times. The optical spectral index ($\alpha$) showed positive correlations with time and R-band magnitude. The weighted mean value of $\alpha$ is 0.94 $\pm$ 0.033 over this STV timescale. \\ 
\\
Studies of flux variability on diverse timescales is a powerful method to understand the radiation mechanisms of blazars: they can provide information about the size, location, and dynamics of the emitting regions \citep{2003A&A...400..487C}. In blazars, the thermal radiation from the accretion disk is generally overwhelmed by the Doppler-boosted nonthermal radiation from the relativistic jet, so the variability on any measurable timescale is most likely explained by the relativistic jet based models. On IDV timescales, blazar variability can reasonably be explained by the shock-in-jet model \citep[e.g.,][]{2015MNRAS.450..541A}. However, in the low states of blazars, the variability may be explained by hotspots on, or instabilities in, the accretion discs \citep[e.g.,][]{1993ApJ...411..602C,1993ApJ...406..420M}. Different models can explain the presence of IDV/STV in the optical (and other) band based on the turbulence behind the shock, or other irregularities in the jet flow produced by variations in the outflow parameters \citep[e.g.,][]{2014ApJ...780...87M,2015JApA...36..255C}. Different optical IDV behaviours have been observed in the LBLs and HBLs subclasses of blazars, with HBLs being relatively less variable in optical bands on IDV time-scales than in X-rays and $\gamma$-rays \citep[e.g.,][]{1996A&A...305...42H,2011MNRAS.416..101G,2016MNRAS.462.1508G}.  The presence of strong magnetic fields in the relativistic jet could be responsible for the different optical microvariability behaviours of LBLs and HBLs \citep{1999A&AS..135..477R}. Stronger magnetic fields in HBLs might interrupt the development of small fluctuation by Kelvin Helmholtz instabilities in jets which could interact with the shocks in jets to produce microvariability if the field strength is greater than the critical value B$_{c}$ given by \citep{2005ChJAS...5..110R}.
\begin{equation}
B_c = \big[4\pi n_e m_e c^2(\gamma^2 - 1)\big]^{1/2} \gamma^{-1},
\end{equation}
where $n_e$ is the local electron density, $m_e$ is the rest mass of electron, and $\gamma$ is the bulk Lorentz factor of the flow. \\
\\
Colour or spectral behaviour are very useful for understanding the emission mechanism of blazars. As mentioned earlier,  blazars generally show one of two different colour behaviours, namely bluer-when-brighter (BWB) and redder-when-brighter (RWB). But in some cases, no colour trend behaviour is seen \citep{2009ApJ...694..174B}.  
In BL Lacs, the BWB trend is predominantly found, whereas the FSRQs usually follow the RWB trend \citep[e.g.,][]{2012AJ....143...23G,2015MNRAS.452.4263G}. Synchrotron models dominated by  one component can explain the BWB behaviour if   the energy distribution of injected fresh electrons, which cause an increase in flux, is harder \citep{1998A&A...333..452K}. When PG 1553$+$113 exhibited IDV during our observations, it has nearly always follows the BWB trend meaning that the increasing flux can be explained by jet synchrotron emission which indicates that particle acceleration efficiency is enhanced \citep{2015MNRAS.451.3882A}. \\ 
\\
From Figure \ref{fig:sed3} we see that as the magnitude increases (or this source dims) the spectrum becomes steeper, and this equivalent way of looking at the BWB trend can also be explained by invoking two components: one is a stable component ($\alpha_{\rm{constant}}$) and the other  contributor to the overall optical emission is a variable component with a flatter slope ($\alpha_{1}$). Blazars show chromatic behaviours when the variable component dominates over the stable component. On short timescales, optical variations can be interpreted as arising from strong chromatic components while long-term optical  variations can be interpreted in terms of a mildly chromatic component \citep{2004A&A...424..497V}. But sometimes, when different pairs of filters are used for colour index measurements, the same source can simultaneously show both BWB and RWB trends \citep{2011MNRAS.418.1640W}. 

\section*{Acknowledgements}
\noindent
We thankfully acknowledge the anonymous reviewer for useful comments.
ACG is partially supported by Chinese Academy of Sciences (CAS) President's International Fellowship Initiative (PIFI) (grant no. 2016VMB073).
SOK acknowledges financial support by the Shota Rustaveli NSF of Georgia under contract PHDF-18-354. 
This research was partially supported by the Bulgarian National Science Fund of the Ministry of Education and Science under grants: KP-06-H28/3 (2018), KP-06-H38/4 (2019) and KP-06-KITAJ/2 (2020). The Skinakas Observatory is a collaborative project of the University of Crete, the Foundation for Research and Technology -- Hellas, and the Max-Planck-Institut f\"ur Extraterrestrische Physik. This research was supported by the Ministry of Education, Science and Technological Development of the Republic of Serbia (contract No.\ 451-03-68/2022-14/200002). GD acknowledges observing grant support from the Institute of Astronomy and Rozhen NAO BAS through the bilateral joint research project ``Gaia Celestial Reference Frame (CRF) and fast variable astronomical objects" (2020-2022, head -- G.\ Damljanovic). MFG acknowledges support from the National Science Foundation of China (grant 11873073), Shanghai Pilot Program for Basic Research Chinese Academy of Science, Shanghai Branch (JCYJ-SHFY2021-013), and the science research grants from the China Manned Space Project with NO. CMSCSST-2021-A06. B.V. is funded by the Swedish Research Council (Vetenskapsr\aa det, grant no. 2017-06372) and is also supported by the The L’Or\'{e}al - UNESCO For Women in Science Sweden Prize with support of the Young Academy of Sweden. She is also supported by the  L’Or\'{e}al - UNESCO For Women in Science International Rising Talents prize 2022. Nordita is partially supported by Nordforsk. HGX is supported by the Ministry of Science and Technology of China (grants No. 2020SKA0110200 and 2018YFA0404601) and the National Science Foundation of China (grants No. 12233005, 11835009, and 11973033). ZZ is thankful for support from the National Key R\&D Program of China (grant No. 2018YFA0404602).

\section*{DATA AVAILABILITY}
\noindent
The data in this article will be shared after one year of the publication of the paper at the reasonable request of the corresponding author.

\appendix
\section{OBSERVATION LOG}
\label{appendix:A1}

\noindent
\begin{table*}
\caption{Observation log for PG 1553$+$113.}            
\label{tab:obs_log}                   
\centering     
\begin{tabular}{c c c c c c c c c}           
\hline                		 
Observation date & Telescope  & Data points &Observation date & Telescope  & Data points &Observation date & Telescope  & Data points \\
yyyy-mm-dd  	&    	&  B, ~V, ~R, ~I & yyyy-mm-dd &    	&  B, ~V, ~R, ~I &	yyyy-mm-dd 	&    	&  B, ~V, ~R, ~I 	\\		 
\hline                          
2019-02-28 	& R1 	&  2, 2, 2, 2 &	2019-04-14	& BS	 	&  0, 0,	43, 	 0 	&	2019-06-05 	&  G   	&  0, 0,  5,  0 	\\
2019-03-04 	& R1	&  2, 2, 2, 2 &	2019-04-14  	&  G   	&  0, 0, 4, 0 	&	2019-06-06  	&  G  	&  0, 0,  4,  0 	\\
2019-03-13 	&  G  	&  0, 0, 5, 0 &	2019-04-15 	& A1 	&  1,31, 31, 31 	& 	2019-06-06  	&AS2 	&  1, 1,  1,  1 	\\ 
2019-03-15	& A2	&  1,12,12, 1 &	2019-04-17	& BS 	&  0, 0, 48,  0	& 	2019-06-08  	&AS2 	&  1, 2,  2,  2 	\\
2019-03-15  &  G	&  0, 0, 5, 0 & 2019-04-18	& BS		&  0, 0,103,  0 	&	2019-06-08  	&  G  	&  0, 0,  6,  0	\\
2019-03-16 	&  G  	&  0, 0, 7, 0 &	2019-04-21	& BS		&  0, 0, 31,  0 	&	2019-06-09  	&  G  	&  0, 0,  4,  0 	\\ 
2019-03-17 	&  G 	&  0, 0, 3, 0 &	2019-04-24 	&  G   	&  0, 0,  3,  0 	&	2019-06-10  	&  G   	&  0, 0,  4,  0 	\\
2019-03-19  & A1	&  1,26,26, 1 &	2019-04-25 	&  G   	&  0, 0,  4,  0 	&	2019-06-13	& A2		&  1, 2,124,  1 	\\
2019-03-20	& A1	&  0,20,20, 0 &	2019-04-26 	&  G   	&  0, 0,  4,  0 	&	2019-06-14	& A2		&  1, 2, 82,  1 	\\
2019-03-21	& A1 	&  1,30,30, 1 &	2019-04-26	& A2		&  1,32, 32,  0 	&	2019-06-20  	&  G   	&  0, 0,  3,  0 	\\
2019-03-22	& A2	&  4,31,31, 0 &	2019-04-27	& A2		&  1,30, 30, 30 	&	2019-06-21  	&  G   	&  0, 0,  4,  0 	\\
2019-03-24	& A1 	&  1, 1, 0, 0 &	2019-04-27	&  G   	&  0, 0,  3,  0 	&	2019-06-22  	&  G   	&  0, 0,  4,  0 	\\
2019-03-25	& A1	&  1,24,24, 1 &	2019-04-28	&  G   	&  0, 0,  6,  0 	&	2019-06-23  	&  G   	&  0, 0,  2,  0 	\\
2019-03-25  &  G  	&  0, 0, 8, 0 &	2019-04-28	& BS		&  0, 0, 35,  0 	&	2019-07-01  	&  G   	&  0, 0,  2,  0 	\\
2019-03-26  &  G  	&  0, 0, 5, 0 &	2019-04-29	& BS		&  0, 0, 54,  0 	&	2019-07-03  	&  G   	&  0, 0,  1,  0 	\\
2019-03-27  &  G  	&  0, 0,10, 0 &	2019-04-29	& R1  	&  2, 2,  2,  2 	&	2019-07-04  	&  G   	&  0, 0,  4,  0 	\\
2019-03-28  &  B  	&  0,15,15,15 &	2019-05-01	& BS		&  0, 0, 32,  0 	& 	2019-07-06  	&AS1 	&  2, 2,  2,  2 	\\
2019-03-29  &  G 	&  0, 0, 5, 0 &	2019-05-02	& R1		&  2, 2,  2,  2 	&	2019-07-17  	&  G  	&  0, 0,  3,  0 	\\
2019-03-29  & R1  	&  3, 3, 3, 3 &	2019-05-02	& R2		&  2, 2,  2,  2	&	2019-07-26  	& R2		&  2, 2,  2,  2 	\\
2019-03-30  &AS2 	&  1, 3, 2, 2 &	2019-05-05	& BS		&  0, 0, 29,  0 	&	2019-07-29  	&  G   	&  0, 0,  4,  0 	\\
2019-03-30  	&  G   	&  0, 0, 5, 0 &	2019-05-06	&  G   	&  0, 0,  5,  0 	&	2019-07-31  	&  G   	&  0, 0,  1,  0 	\\
2019-03-31  	&AS2 	&  1, 1, 1, 1 &	2019-05-07	& A1		&  2, 2,  2,  2 	&	2019-08-01  	&  G   	&  0, 0,  4,  0 	\\
2019-03-31  	&  G   	&  0, 0, 5, 0 &	2019-05-07	&  G   	&  0, 0,  2,  0 	&	2019-08-02  	&  G   	&  0, 0,  4,  0 	\\
2019-04-02	& A1	 	&  2,37,37, 2 &	2019-05-08	&  G   	&  0, 0,  4,  0 	&	2019-08-08  	& R1		&  2, 2,  2,  2 	\\
2019-04-06  	&AS2 	&  1, 1, 1, 1 &	2019-05-10	&  G   	&  0, 0,  4,  0 	& 	2019-08-10  	&  S 	&  0, 3,  3,  3 	\\
2019-04-07  	&  G   	&  0, 0, 5, 0 &	2019-05-13	&  G   	&  0, 0,  4,  0 	& 	2019-08-11  	&  S		&  3, 3,  3,  3 	\\ 
2019-04-08  	&  G   	&  0, 0, 5, 0 &	2019-05-14	& A1		&  1, 1,  1,  1 	&	2019-08-12  	&  S 	&  3, 3,  3,  3 	\\
2019-04-08	& A1	  	&  1,41,41,41 &	2019-05-15	& A2		&  1,36, 36, 36 	&	2019-08-13  	&  S 	&  3, 3,  3,  3 	\\
2019-04-09  	&  G   	&  0, 0, 6, 0 &	2019-05-15	&  G   	&  0, 0,  4,  0 	&	2019-08-14  	&  S 	&  3, 3,  3,  3 	\\
2019-04-09	& A1	  	&  2, 2, 2, 2 &	2019-05-18	& A1		&  0, 0,  1,  1	&	2019-08-14  	&  G   	&  0, 0,  2,  0 	\\

2019-04-10  	&  G   	&  0, 0, 8, 0 &	2019-05-21	&  G   	&  0, 0,  4,  0 	&	2019-08-15  	&  G   	&  0, 0,  3,  0 	\\ 
2019-04-10	& A1	  	&  1,37,37,37 &	2019-05-23	&  G   	&  0, 0,  3,  0 	&	2019-08-18  	&  G   	&  0, 0,  3,  0 	\\
2019-04-11  	&  G   	&  0, 0, 4, 0 &	2019-05-25	&  G   	&  0, 0,  4,  0 	&	2019-08-28  	&AS1 	&  2, 2,  2,  2 	\\ 
2019-04-11	& A1	  	&  2,32,32,30 &	2019-05-25	& A1		&  2, 1,  1,  1 	&	2019-08-31  	&AS1 	&  2, 2,  2,  2 	\\
2019-04-12	& A1	  	&  2, 2, 2, 2 &	2019-05-26	& A1		&  1, 2,  2,  1 	&	2019-09-03  	& R1   	&  2, 2,  2,  2 	\\
2019-04-12  	&  G   	&  0, 0, 8, 0 &	2019-05-26	&  G   	&  0, 0,  4,  0 	&	2019-09-12  	&  G  	&  0, 0,  3,  0 	\\
2019-04-13  	&  G   	&  0, 0, 4, 0 &	2019-05-30	& A2		&  1,82, 82,  1 	&	2019-09-29  	&  G		&  0, 0,  3,  0 	\\
2019-04-13	& A1 	&  2, 2, 2, 2 &	2019-06-04	& A2		&  1,34, 34,  1 	&	2019-10-22  	&  G   	&  0, 0,  5,  0 	\\ 
2019-04-13	& BS 	&  0, 0,29, 0 &	2019-06-04 	&  G   	&  0, 0,  5,  0 	&	2019-11-08  	&  G   	&  0, 0,  5,  0 	\\\hline  \end{tabular}

\end{table*}

\section{RESULTS OF IDV}

\begin{table*}
\caption{Results of IDV analysis of PG 1553$+$113}            
\label{appendix:B1}
\centering 
\begin{tabular}{lcccccccccc} \hline \hline                		 
Observation date & Band & \multicolumn{3}{c}{{\it Power-enhanced  F-test}}  & \multicolumn{3}{c}{{\it Nested ANOVA}} &Status &   Amplitude\\
\cmidrule[0.03cm](r){3-5}\cmidrule[0.03cm](r){6-8} yyyy-mm-dd & &DoF($\nu_1$,$\nu_2$ ) & $F_{enh}$ & $F_c$  & DoF($\nu_1$,$\nu_2$ ) & $F$ & $F_c$& & $\%$& \\\hline
	 	  
20190315  & V     & 11, 22 & 1.42 & 3.18 & 2, 9 & 1.98 & 8.02 & NV & --  \\ 
	  	  & R     & 11, 22 & 1.70 & 3.18 & 2, 9 & 2.33 & 8.02 & NV & --  \\ 
	      & V$-$R   & 11, 22 & 1.47 & 3.18 & 2, 9 & 0.91 & 8.02 & NV & --   \\

20190319  & V     & 25, 50 & 0.64 & 2.16    & 4, 20 & 0.87 & 4.43 & NV & --  \\ 
	  	  & R     & 25, 50 & 0.84 & 2.16    & 4, 20 & 2.83 & 4.43 & NV & --  \\ 
	      &V$-$R    & 25, 50 & 0.64 & 2.16    & 4, 20 & 0.94 & 4.43 & NV & --  \\ 
	      
20190320  & V     & 19, 38 & 1.32 & 2.42    & 3, 16 & 0.31 & 5.29 & NV & --  \\ 
	  	  & R     & 19, 38 & 1.42 & 2.42    & 3, 16 & 1.76 & 5.29 & NV & --  \\ 
	      &V$-$R    & 19, 38 & 0.72 & 2.42    & 3, 16 & 0.81 & 5.29 & NV & --  \\
	      
20190321  & V     & 29, 58 & 1.89 & 2.05    & 5, 24 & 1.43 & 3.89 & NV & --  \\ 
	  	  & R     & 29, 58 & 4.08 & 2.05    & 5, 24 & 2.38 & 3.89 & NV & --  \\ 
	      & V$-$R   & 29, 58 & 1.66 & 2.05    & 5, 24 & 1.02 & 3.89 & NV & --  \\
	      
20190322  & V     & 30, 60 & 1.82 & 2.03    & 5, 24 & 0.52 & 3.89 & NV & --  \\ 
	  	  & R     & 30, 60 & 0.48 & 2.03    & 5, 24 & 0.23 & 3.89 & NV & --  \\ 
	      & V$-$R   & 30, 60 & 1.15 & 2.03    & 5, 24 & 0.38 & 3.89 & NV & --  \\
	      
20190325  & V     & 23, 46 & 1.62 & 2.24    & 5, 18 & 2.56 & 4.25 & NV & --  \\ 
	      & R     & 23, 46 & 1.34 & 2.24    & 5, 18 & 1.33 & 4.25 & NV & --  \\ 
	      & V$-$R   & 23, 46 & 1.87 & 2.24    & 5, 18 & 2.32 & 4.25 & NV & --  \\ 
	      
20190402  & V     & 36, 72 & 0.74 & 1.91    & 6, 28 & 0.86 & 3.52 & NV & --  \\ 
	      & R     & 36, 72 & 1.52 & 1.91    & 6, 28 & 7.83 & 3.52 & NV & --  \\ 
		  & V$-$R   & 36, 72 & 1.07 & 1.91    & 6, 28 & 5.65 & 3.52 & NV & --  \\ 
		  
20190408  & V     & 40, 80 & 2.55 & 1.85    & 7, 32 & 2.57 & 3.25 & NV & --  \\ 
	      & R     & 40, 80 & 2.18 & 1.85    & 7, 32 & 0.67 & 3.25 & NV & --  \\ 
	  	  & I     & 40, 80 & 1.15 & 1.85    & 7, 32 & 1.35 & 3.25 & NV & -- \\
	  	  & V$-$R   & 40, 80 & 1.28 & 1.85    & 7, 32 & 2.22 & 3.25 & NV & --  \\
	  	  
20190410  & V     & 37, 74 & 0.99 & 1.89    & 6, 28 & 0.62 & 3.52 & NV & --  \\ 
	      & R     & 37, 74 & 1.22 & 1.89    & 6, 28 & 0.62 & 3.52 & NV & --  \\ 
	  	  & I     & 37, 74 & 0.84 & 1.89    & 6, 28 & 1.01 & 3.52 & NV & --  \\
	      & V$-$R   & 37, 74 & 0.69 & 1.89    & 6, 28 & 0.93 & 3.52 & NV & --  \\
	      
20190411  & V     & 31, 62 & 0.33 & 2.01    & 5, 24 & 1.58 & 3.89 & NV &  -- \\ 
	      & R     & 31, 62 & 0.42 & 2.01    & 5, 24 & 1.53 & 3.89 & NV &  -- \\ 
	  	  & I     & 29, 58 & 0.49 & 2.03    & 5, 24 & 1.15 & 3.89 & NV &  -- \\
	      & V$-$R   & 31, 62 & 0.26 & 2.01    & 5, 24 & 2.39 & 3.89 & NV &  -- \\
	      
20190413  & R     & 30, 60 & 0.97 & 2.03    & 5, 24 & 2.65 & 3.89 & NV &  -- \\ 

20190414  & R     & 42, 84 & 3.66 & 1.82    & 7, 32 & 1.13 & 3.26 & NV &  -- \\

20190415  & V     & 30, 60 & 1.32 & 2.028   & 5, 24 & 5.77 & 3.89 & NV &  -- \\
		  & R     & 30, 60 & 1.62 & 2.028   & 5, 24 & 9.87 & 3.89 & NV &   -- \\
		  & I     & 30, 60 & 1.54 & 2.028   & 5, 24 & 1.46 & 3.89 & NV &  -- \\ 
	      & V$-$R   & 30, 60 & 1.48 & 2.028   & 5, 24 & 2.31 & 3.89 & NV &  -- \\ 
	      
20190417  & R     & 47, 94 & 2.78 & 1.76    & 11, 36 & 9.44 & 2.79 &  V & 14.98  \\

20190418  & R     & 102,204& 2.93 & 1.48    & 19,80  & 4.19 & 2.14 &  V & 8.89  \\

20190421  & R     & 30, 60 & 3.81 & 2.03    &  5, 24 & 2.37 & 3.89 & NV & --  \\

20190426  & V     & 31, 62 & 0.99 & 2.01    & 5,24 & 0.43 & 3.89 & NV &  -- \\
		  & R     & 31, 62 & 1.02 & 2.01    & 5,24 & 0.46 & 3.89 & NV &  -- \\
		  & V$-$R   & 31, 62 & 1.23 & 2.01    & 5,24 & 0.35 & 3.89 & NV &  -- \\
20190427  & V     & 29, 58 & 0.86 & 2.05    & 5, 24 & 3.20 & 3.89 & NV & --  \\
		  & R     & 29, 58 & 1.04 & 2.05    & 5, 24 & 1.70 & 3.89 & NV & -- \\
		  & I     & 29, 58 & 0.71 & 2.05    & 5, 24 & 2.30 & 3.89 & NV & --  \\ 
		  & V$-$R   & 29, 58 & 1.37 & 2.05    & 5, 24 & 3.49 & 3.89 & NV & -- \\
		  
20190428  & R     & 34, 68 & 1.25 & 1.95    & 6,28 & 3.58  & 3.52 & NV & --  \\

20190429  & R     & 53, 106& 0.56 & 1.71    & 9,40 & 3.91  & 2.89 & NV & --  \\

20190501  & R     & 31, 62 & 0.59 & 2.01    & 5,24 & 1.99  & 3.89 & NV & --  \\

20190505  & R     & 28, 56 & 0.92 & 2.08    & 4, 20 & 2.31 & 4.43 & NV & --  \\

20190515  & V     & 35, 70 & 6.53 & 1.93    & 6, 28 & 10.36 & 3.52 &  V & 11.82  \\
		  & R     & 35, 70 & 3.13 & 1.93    & 6, 28 &  8.57 & 3.52 &  V & 9.02  \\
		  & I     & 35, 70 & 3.21 & 1.93    & 6, 28 &  3.75 & 3.52 &  V & 8.19  \\
		  & V$-$R   & 35, 70 & 0.82 & 1.93    & 6, 28 &  1.44 & 3.52 & NV & --  \\

20190530  & V     & 81 ,162& 1.29 & 1.55    & 15,64 & 1.25 & 2.33 & NV & --  \\ 
	      & R     & 81 ,162& 1.26 & 1.55    & 15,64 & 1.89 & 2.33 & NV & --  \\
	      & V$-$R   & 81, 162& 0.82 & 1.55    & 15,64 & 1.36 & 2.33 & NV & --  \\
	      
20190604  & V     & 33, 66 & 1.88 & 1.96    & 5, 24 & 1.01 & 3.89 & NV & --  \\ 
	      & R     & 33, 66 & 0.75 & 1.96    & 5, 24 & 0.94 & 3.89 & NV & --  \\
		  & V$-$R   & 33, 66 & 0.91 & 1.96    & 5, 24 & 1.04 & 3.89 & NV & --  \\
20190613  & R     &123, 246& 1.97 & 1.43    & 23,96 & 2.65 & 2.00 &  V &  3.91 \\
20190614  & R     & 81, 162& 0.47 & 1.55    & 15,64 & 1.57 & 2.33 & NV & --  \\\hline                          
\end{tabular}

\end{table*}


\begin{thebibliography}{}

\bibitem[Ackermann et al.(2015)]{2015ApJ...813L..41A}
Ackermann, M., Ajello, M., Albert, A., et al.\ 2015, \apjl, 813, L41

\bibitem[Agarwal \& Gupta(2015)]{2015MNRAS.450..541A} Agarwal, A. \& Gupta, A.~C.\ 2015, \mnras, 450, 541. 

\bibitem[Agarwal et al.(2015)]{2015MNRAS.451.3882A} Agarwal, A., Gupta, A.~C., Bachev, R., et al.\ 2015, \mnras, 451, 3882.

\bibitem[Agarwal et al.(2016)]{2016MNRAS.455..680A} Agarwal, A., Gupta, A.~C., Bachev, R., et al.\ 2016, \mnras, 455, 680. 

\bibitem[Agarwal et al.(2019)]{2019MNRAS.488.4093A} Agarwal, A., Cellone, S.~A., Andruchow, I., et al.\ 2019, \mnras, 488, 4093. 

\bibitem[Agarwal et al.(2021)]{2021A&A...645A.137A} Agarwal, A., Mihov, B., Andruchow, I., et al.\ 2021, \aap, 645, A137. 

\bibitem[Aharonian et al.(2006)]{2006A&A...448L..19A}
Aharonian, F., Akhperjanian, A.~G., Bazer-Bachi, A.~R., et al.\ 2006, \aap, 448, L19

\bibitem[Andruchow et al.(2011)]{2011A&A...531A..38A} Andruchow, I., Combi, J.~A., Mu{\~n}oz-Arjonilla, A.~J., et al.\ 2011, \aap, 531, A38. 

\bibitem[Bachev et al.(2012)]{2012MNRAS.424.2625B} Bachev, R., Semkov, E., Strigachev, A., et al.\ 2012, \mnras, 424, 2625. 

\bibitem[Bachev et al.(2017)]{2017MNRAS.471.2216B} Bachev, R., Popov, V., Strigachev, A., et al.\ 2017, \mnras, 471, 2216. 

\bibitem[Bai et al.(1998)]{1998A&AS..132...83B} Bai, J.~M., Xie, G.~Z., Li, K.~H., et al.\ 1998, \aaps, 132, 83. 

\bibitem[B{\"o}ttcher et al.(2009)]{2009ApJ...694..174B} B{\"o}ttcher, M., Fultz, K., Aller, H.~D., et al.\ 2009, \apj, 694, 174. 

\bibitem[Calafut \& Wiita(2015)]{2015JApA...36..255C} Calafut, V. \& Wiita, P.~J.\ 2015, Journal of Astrophysics and Astronomy, 36, 255.

\bibitem[Carini et al.(1990)]{1990AJ....100..347C} Carini, M.~T., Miller, H.~R., \& Goodrich, B.~D.\ 1990, \aj, 100, 347. 

\bibitem[Carini et al.(1992)]{1992AJ....104...15C} Carini, M.~T., Miller, H.~R., Noble, J.~C., et al.\ 1992, \aj, 104, 15. 

\bibitem[Chakrabarti \& Wiita(1993)]{1993ApJ...411..602C}
Chakrabarti, S.~K., \& Wiita, P.~J.\ 1993, \apj, 411, 602

\bibitem[Ciprini et al.(2003)]{2003A&A...400..487C}
Ciprini, S., Tosti, G., Raiteri, C.~M., et al.\ 2003, \aap, 400, 487

\bibitem[Clements \& Carini(2001)]{2001AJ....121...90C} Clements, S.~D. \& Carini, M.~T.\ 2001, \aj, 121, 90. 

\bibitem[Covino et al.(2020)]{2020ApJ...895..122C} Covino, S., Landoni, M., Sandrinelli, A., et al.\ 2020, \apj, 895, 122. 

\bibitem[de Diego et al.(1998)]{1998ApJ...501...69D} de Diego, J.~A., Dultzin-Hacyan, D., Ram{\'\i}rez, A., et al.\ 1998, \apj, 501, 69. 

\bibitem[de Diego(2014)]{2014AJ....148...93D} de Diego, J.~A.\ 2014, \aj, 148, 93. 

\bibitem[de Diego et al.(2015)]{2015AJ....150...44D} de Diego, J.~A., Polednikova, J., Bongiovanni, A., et al.\ 2015, \aj, 150, 44. 

\bibitem[Dhiman et al.(2021)]{2021MNRAS.506.1198D} Dhiman, V., Gupta, A.~C., Gaur, H., et al.\ 2021, \mnras, 506, 1198. 

\bibitem[{{Edelson} \& {Krolik}(1988)}]{1988ApJ...333..646E}
{Edelson}, R.~A., \& {Krolik}, J.~H. 1988, \apj, 333, 646

\bibitem[{{Falomo} \& {Treves}(1990)}]{1990PASP..102.1120F}
{Falomo}, R., \& {Treves}, A. 1990, \pasp, 102, 1120

\bibitem[Falomo et al.(1994)]{1994ApJS...93..125F} Falomo, R., Scarpa, R., \& Bersanelli, M.\ 1994, \apjs, 93, 125. 

\bibitem[Fossati et al.(1998)]{1998MNRAS.299..433F}
Fossati, G., Maraschi, L., Celotti, A., et al.\ 1998, \mnras, 299, 433

\bibitem[Fan et al.(1997)]{1997A&AS..125..525F} Fan, J.~H., Xie, G.~Z., Lin, R.~G., et al.\ 1997, \aaps, 125, 525. 

\bibitem[Fan et al.(1998)]{1998ApJ...507..173F} Fan, J.~H., Xie, G.~Z., Pecontal, E., et al.\ 1998, \apj, 507, 173. 

\bibitem[Fan et al.(2001)]{2001A&A...369..758F} Fan, J.~H., Qian, B.~C., \& Tao, J.\ 2001, \aap, 369, 758. 


\bibitem[Gaur et al.(2012a)]{2012AJ....143...23G} Gaur, H., Gupta, A.~C., \& Wiita, P.~J.\ 2012, \aj, 143, 23.

\bibitem[Gaur et al.(2012b)]{2012MNRAS.420.3147G} Gaur, H., Gupta, A.~C., Strigachev, A., et al.\ 2012, \mnras, 420, 3147. 

\bibitem[Gaur et al.(2012c)]{2012MNRAS.425.3002G} Gaur, H., Gupta, A.~C., Strigachev, A., et al.\ 2012, \mnras, 425, 3002. 

\bibitem[Gaur et al.(2015)]{2015MNRAS.452.4263G} Gaur, H., Gupta, A.~C., Bachev, R., et al.\ 2015, \mnras, 452, 4263. 

\bibitem[Gaur et al.(2019)]{2019MNRAS.484.5633G} Gaur, H., Gupta, A.~C., Bachev, R., et al.\ 2019, \mnras, 484, 5633. 

\bibitem[Ghosh et al.(2000)]{2000ApJ...537..638G} Ghosh, K.~K., Ramsey, B.~D., Sadun, A.~C., et al.\ 2000, \apj, 537, 638. 

\bibitem[Gopal-Krishna et al.(1993)]{1993MNRAS.262..963G} Gopal-Krishna, Sagar, R., \& Wiita, P.~J.\ 1993, \mnras, 262, 963. 

\bibitem[Gopal-Krishna et al.(2011)]{2011MNRAS.416..101G} Gopal-Krishna, Goyal, A., Joshi, S., et al.\ 2011, \mnras, 416, 101. 

\bibitem[Goyal et al.(2013)]{2013JApA...34..273G} Goyal, A., Mhaskey, M., Gopal-Krishna, et al.\ 2013, Journal of Astrophysics and Astronomy, 34, 273. 

\bibitem[{{Green} {et~al.}(1986){Green}, {Schmidt}, \& {Liebert}}]{1986ApJS...61..305G}
{Green}, R.~F., {Schmidt}, M., \& {Liebert}, J. 1986, \apjs, 61, 305

\bibitem[{{Gupta} {et~al.}(2004){Gupta}, {Banerjee}, {Ashok}, \& {Joshi}}]{2004A&A...422..505G}
{Gupta}, A.~C., {Banerjee}, D.~P.~K., {Ashok}, N.~M., \& {Joshi}, U.~C. 2004, \aap, 422, 505

\bibitem[Gupta \& Joshi(2005)]{2005A&A...440..855G} Gupta, A.~C. \& Joshi, U.~C.\ 2005, \aap, 440, 855. 

\bibitem[Gupta et al.(2008)]{2008AJ....135.1384G} Gupta, A.~C., Fan, J.~H., Bai, J.~M., et al.\ 2008, \aj, 135, 1384. 

\bibitem[Gupta et al.(2016a)]{2016MNRAS.458.1127G} Gupta, A.~C., Agarwal, A., Bhagwan, J., et al.\ 2016, \mnras, 458, 1127. 

\bibitem[Gupta et al.(2016b)]{2016MNRAS.462.1508G} Gupta, A.~C., Kalita, N., Gaur, H., et al.\ 2016, \mnras, 462, 1508. 

\bibitem[Gupta et al.(2017)]{2017MNRAS.465.4423G} Gupta, A.~C., Agarwal, A., Mishra, A., et al.\ 2017, \mnras, 465, 4423. 

\bibitem[Gupta et al.(2019)]{2019AJ....157...95G} Gupta, A.~C., Gaur, H., Wiita, P.~J., et al.\ 2019, \aj, 157, 95. 

\bibitem[Gu et al.(2006)]{2006A&A...450...39G} Gu, M.~F., Lee, C.-U., Pak, S., et al.\ 2006, \aap, 450, 39. 

\bibitem[Heidt \& Wagner(1996)]{1996A&A...305...42H} Heidt, J. \& Wagner, S.~J.\ 1996, \aap, 305, 42

\bibitem[Hu et al.(2006)]{2006MNRAS.371.1243H} Hu, S.~M., Zhao, G., Guo, H.~Y., et al.\ 2006, \mnras, 371, 1243. 

\bibitem[Hufnagel \& Bregman(1992)]{1992ApJ...386..473H} 
Hufnagel, B.~R., \& Bregman, J.~N.\ 1992, \apj, 386, 473

\bibitem[Jannuzi et al.(1994)]{1994ApJ...428..130J} Jannuzi, B.~T., Smith, P.~S., \& Elston, R.\ 1994, \apj, 428, 130. 

\bibitem[Johnson et al.(2019)]{2019ApJ...884L..31J} Johnson, S.~D., Mulchaey, J.~S., Chen, H.-W., et al.\ 2019, \apjl, 884, L31. 

\bibitem[Joshi et al.(2011)]{2011MNRAS.412.2717J} Joshi, R., Chand, H., Gupta, A.~C., et al.\ 2011, \mnras, 412, 2717. 

\bibitem[Kirk et al.(1998)]{1998A&A...333..452K} Kirk, J.~G., Rieger, F.~M., \& Mastichiadis, A.\ 1998, \aap, 333, 452

\bibitem[Kshama et al.(2017)]{2017MNRAS.466.2679K} Kshama, S.~K., Paliya, V.~S., \& Stalin, C.~S.\ 2017, \mnras, 466, 2679. 

\bibitem[Mangalam \& Wiita(1993)]{1993ApJ...406..420M} 
Mangalam, A.~V., \& Wiita, P.~J.\ 1993, \apj, 406, 420

\bibitem[Marscher(2014)]{2014ApJ...780...87M} Marscher, A.~P.\ 2014, \apj, 780, 87

\bibitem[Massaro et al.(1998)]{1998MNRAS.299...47M} Massaro, E., Nesci, R., Maesano, M., et al.\ 1998, \mnras, 299, 47. 

\bibitem[Miller \& Green(1983)]{1983BAAS...15..957M} Miller, H.~R., \& Green, R.~F.\ 1983, \baas, 15, 957

\bibitem[Miller et al.(1989)]{1989Natur.337..627M} Miller, H.~R., Carini, M.~T., \& Goodrich, B.~D.\ 1989, \nat, 337, 627

\bibitem[Padovani \& Giommi(1995)]{1995MNRAS.277.1477P} Padovani, P., \& Giommi, P.\ 1995, \mnras, 277, 1477

\bibitem[Pandey et al.(2017)]{2017ApJ...841..123P} Pandey, A., Gupta, A.~C., \& Wiita, P.~J.\ 2017, \apj, 841, 123. 

\bibitem[Pandey et al.(2019)]{2019ApJ...871..192P} Pandey, A., Gupta, A.~C., Wiita, P.~J., \& Tiwari, S.~N.\ 2019, \apj, 871, 192

\bibitem[Pandey et al.(2020a)]{2020ApJ...890...72P} Pandey, A., Gupta, A.~C., Kurtanidze, S.~O., et al.\ 2020, \apj, 890, 72. 

\bibitem[Pandey et al.(2020b)]{2020MNRAS.496.1430P} Pandey, A., Gupta, A.~C., Damljanovic, G., et al.\ 2020, \mnras, 496, 1430. 

\bibitem[Papadakis et al.(2003)]{2003A&A...397..565P} Papadakis, I.~E., Boumis, P., Samaritakis, V., et al.\ 2003, \aap, 397, 565. 

\bibitem[Papadakis et al.(2007)]{2007A&A...470..857P} Papadakis, I.~E., Villata, M., \& Raiteri, C.~M.\ 2007, \aap, 470, 857. 

\bibitem[Pasierb et al.(2020)]{2020MNRAS.492.1295P} Pasierb, M., Goyal, A., Ostrowski, M., et al.\ 2020, \mnras, 492, 1295. 

\bibitem[Pe{\~n}il et al.(2020)]{2020ApJ...896..134P} Pe{\~n}il, P., Dom{\'\i}nguez, A., Buson, S., et al.\ 2020, \apj, 896, 134. 

\bibitem[Polednikova et al.(2016)]{2016MNRAS.460.3950P} Polednikova, J., Ederoclite, A., de Diego, J.~A., et al.\ 2016, \mnras, 460, 3950. 

\bibitem[Prokhorov \& Moraghan(2017)]{2017MNRAS.471.3036P} Prokhorov, D.~A. \& Moraghan, A.\ 2017, \mnras, 471, 3036. 

\bibitem[Raiteri et al.(1998)]{1998A&AS..127..445R} Raiteri, C.~M., Ghisellini, G., Villata, M., et al.\ 1998, \aaps, 127, 445. 

\bibitem[Raiteri et al.(2001)]{2001A&A...377..396R} Raiteri, C.~M., Villata, M., Aller, H.~D., et al.\ 2001, \aap, 377, 396. 

\bibitem[Raiteri et al.(2007)]{2007A&A...464..871R} Raiteri, C.~M., Villata, M., Capetti, A., et al.\ 2007, \aap, 464, 871. 

\bibitem[Raiteri et al.(2015)]{2015MNRAS.454..353R} Raiteri, C.~M., Stamerra, A., Villata, M., et al.\ 2015, \mnras, 454, 353. 

\bibitem[Rani et al.(2010)]{2010MNRAS.404.1992R} Rani, B., Gupta, A.~C., Strigachev, A., et al.\ 2010, \mnras, 404, 1992. 

\bibitem[Romero et al.(1999)]{1999A&AS..135..477R} 
Romero, G.~E., Cellone, S.~A., \& Combi, J.~A.\ 1999, \aaps, 135, 477

\bibitem[Romero(2005)]{2005ChJAS...5..110R} Romero, G.~E.\ 2005, Chinese Journal of Astronomy and Astrophysics Supplement, 5, 110

\bibitem[Sagar et al.(1996)]{1996MNRAS.281.1267S} Sagar, R., Gopal-Krishna, \& Wiita, P.~J.\ 1996, \mnras, 281, 1267. 

\bibitem[Sandrinelli et al.(2018)]{2018A&A...615A.118S} Sandrinelli, A., Covino, S., Treves, A., et al.\ 2018, \aap, 615, A118. 

\bibitem[Sillanpaa et al.(1996a)]{1996A&A...305L..17S} Sillanpaa, A., Takalo, L.~O., Pursimo, T., et al.\ 1996, \aap, 305, L17

\bibitem[Sillanpaa et al.(1996b)]{1996A&A...315L..13S} Sillanpaa, A., Takalo, L.~O., Pursimo, T., et al.\ 1996, \aap, 315, L13

\bibitem[Stalin et al.(2004)]{2004MNRAS.350..175S} Stalin, C.~S., Gopal-Krishna, Sagar, R., et al.\ 2004, \mnras, 350, 175. 

\bibitem[Stalin et al.(2005)]{2005MNRAS.356..607S} Stalin, C.~S., Gupta, A.~C., Gopal-Krishna, et al.\ 2005, \mnras, 356, 607. 

\bibitem[Stetson(1987)]{1987PASP...99..191S} Stetson, P.~B.\ 1987, \pasp, 99, 191. 

\bibitem[Stetson(1992)]{1992ASPC...25..297S} Stetson, P.~B.\ 1992, Astronomical Data Analysis Software and Systems I, 25, 297

\bibitem[Tavani et al.(2018)]{2018ApJ...854...11T} 
Tavani, M., Cavaliere, A., Munar-Adrover, P., \& Argan, A.\ 2018, \apj, 854, 11

\bibitem[{{Urry} \& {Padovani}(1995)}]{1995PASP..107..803U}
{Urry}, C.~M., \& {Padovani}, P. 1995, \pasp, 107, 803

\bibitem[Villata et al.(2002)]{2002A&A...390..407V} Villata, M., Raiteri, C.~M., Kurtanidze, O.~M., et al.\ 2002, \aap, 390, 407. 

\bibitem[Villata et al.(2004a)]{2004A&A...421..103V} Villata, M., Raiteri, C.~M., Kurtanidze, O.~M., et al.\ 2004, \aap, 421, 103.

\bibitem[Villata et al.(2004b)]{2004A&A...424..497V} Villata, M., Raiteri, C.~M., Aller, H.~D., et al.\ 2004, \aap, 424, 497. 

\bibitem[Villata et al.(2006)]{2006A&A...453..817V} Villata, M., Raiteri, C.~M., Balonek, T.~J., et al.\ 2006, \aap, 453, 817. 

\bibitem[{{Wagner} \& {Witzel}(1995)}]{1995ARA&A..33..163W}
{Wagner}, S.~J., \& {Witzel}, A. 1995, \araa, 33, 163

\bibitem[Wu et al.(2011)]{2011MNRAS.418.1640W} Wu, J., Zhou, X., Ma, J., et al.\ 2011, \mnras, 418, 1640. 

\bibitem[Xie et al.(1992)]{1992ApJS...80..683X} Xie, G.~Z., Li, K.~H., Liu, F.~K., et al.\ 1992, \apjs, 80, 683. 

\bibitem[Xie et al.(1994)]{1994A&AS..106..361X} Xie, G.~Z., Li, K.~H., Zhang, Y.~H., et al.\ 1994, \aaps, 106, 361

\bibitem[Xie et al.(2002)]{2002MNRAS.329..689X} Xie, G.~Z., Zhou, S.~B., Dai, B.~Z., et al.\ 2002, \mnras, 329, 689. 

\end{thebibliography}
\end{document}